\documentclass[prb,twocolumn,preprintnumbers,amsmath,amssymb]{revtex4}

\usepackage{amsmath,amssymb,amsfonts}
\usepackage[dvips]{graphicx}
\usepackage[latin1]{inputenc}
\usepackage{subfigure}

\begin{document}

\title{Static correlation beyond the random phase approximation: Dissociating H$_2$ with the Bethe-Salpeter equation and time-dependent GW}

\author{Thomas Olsen}
\email{tolsen@fysik.dtu.dk}
\author{Kristian S. Thygesen}

\affiliation{Center for Atomic-Scale Materials Design (CAMD) and Center for Nanostructured Graphene (CNG),
	     Department of Physics, Technical University of Denmark,
	     DK--2800 Kongens Lyngby, Denmark}

\date{\today}

\begin{abstract}
We investigate various approximations to the correlation energy of a H$_2$ molecule in the dissociation limit, where the ground state is poorly described by a single Slater determinant. The correlation energies are derived from the density response function and it is shown that response functions derived from Hedin's equations (Random Phase Approximation (RPA), Time-dependent Hartree-Fock (TDHF), Bethe-Salpeter equation (BSE), and Time-Dependent GW (TDGW)) all reproduce the correct dissociation limit. We also show that the BSE improves the correlation energies obtained within RPA and TDHF significantly for intermediate binding distances. A Hubbard model for the dimer allow us to obtain exact analytical results for the various approximations, which is readily compared with the exact diagonalization of the model. Moreover, the model is shown to reproduce all the qualitative results from the \textit{ab initio} calculations and confirms that BSE greatly improves the RPA and TDHF results despite the fact that the BSE excitation spectrum breaks down in the dissociation limit. In contrast, Second Order Screened Exchange (SOSEX) gives a poor description of the dissociation limit, which can be attributed to the fact that it cannot be derived from an irreducible response function.
\end{abstract}

\maketitle

\section{Introduction}
In many-body quantum theory, the presence of two-particle interactions, renders the wavefunctions prohibitly complicated objects, that can only be obtained in simple models or for systems containing very few particles. The Coulomb interaction represents a prominent example in electronic structure problems and a large part of the research in chemical and solid state physics is devoted to developing and applying approximate treatments of this interaction. If one is interested in the energy spectrum of a particular system, the simplest approach is to diagonalize the non-interacting Hamiltonian and correct for the Coulomb interactions perturbatively. However, the Coulomb interaction is by no means weak and the convergence of such a perturbative approach is questionable. In particular, the non-interacting wavefunctions are Slater determinants composed of single particle orbitals and if the true many-body wavefunctions are poorly approximated by such an approximation, one would expect standard perturbation theory to fail.

Another approach to the problem is to apply a mean-field approach like Hartree-Fock (HF) or Kohn-Sham Density Functional Theory (KS-DFT), where the Coulomb interaction is replaced by an average value, which acts as an external potential. As a consequence, the many-body wavefunction becomes a Slater determinant composed of single-particle orbitals that are easily obtained as eigenfunctions of the mean-field Hamiltonian. Again, one faces the problem that if the true many-body wavefunction is poorly approximated by a Slater determinant, the mean field approach is bound to yield a bad approximation for the wavefunctions. Nevertheless, in KS-DFT the mean field Hamiltonian does not have to approximate the interacting system, but it is still possible to calculate the correct ground state energy provided one knows the exact exchange-correlation functional. However, in practice the exchange-correlation functional has to be approximated and will typically perform poorly if the mean field Hamiltonian does not describe the same physics as the interacting Hamiltonian.\cite{helbig}

Systems whose ground states are poorly described by a Slater determinant are referred to as static correlated. The generic example of static correlation is the Hydrogen molecule in the dissociation limit, which is well described by the Heitler-London wavefunction\cite{heitler}
\begin{align}\label{heitler}
\psi_{HL}(\mathbf{r}_1,\mathbf{r}_2)=\big[&\varphi^{1}_\uparrow(\mathbf{r}_1)\varphi^{2}_\downarrow(\mathbf{r}_2)+\varphi^{2}_\uparrow(\mathbf{r}_1)\varphi^{1}_\downarrow(\mathbf{r}_2)\\
-&\varphi^{1}_\downarrow(\mathbf{r}_1)\varphi^{2}_\uparrow(\mathbf{r}_2)-\varphi^{2}_\downarrow(\mathbf{r}_1)\varphi^{1}_\uparrow(\mathbf{r}_2)\big]/2\notag,
\end{align}
where $\varphi^{N}$ denotes a $1s$ orbital of atom $N$. This state correctly captures the correlated effect that if one electron is found on one atom, the other electron will be on the other atom. In contrast, any mean-field approach will yield the Slater determinant
\begin{align}\label{slater}
\psi_{S}(\mathbf{r}_1,\mathbf{r}_2)=\big[\sigma_\uparrow(\mathbf{r}_1)\sigma_\downarrow(\mathbf{r}_2)-\sigma_\downarrow(\mathbf{r}_1)\sigma_\uparrow(\mathbf{r}_2)\big]/\sqrt{2},
\end{align}
where $\sigma(\mathbf{r})$ are bonding $\sigma$-orbitals. In this state, the probability of finding both electrons on the same atom is always $1/2$ independent of the interatomic distance. HF theory then predicts a ground state energy in the dissociation limit, which is $\sim$ 7 eV too high and KS-DFT with standard semi-local functionals, predicts the energy to be $\sim$ 2 eV too high.\cite{olsen_rpa2} The true many-body wavefunction will resemble \eqref{slater} at equilibrium binding distance and approach \eqref{heitler} in the dissociation limit. Which of the two wavefunction provide the better description is determined by the relative magnitudes of the hybridization integrals and the Coulomb repulsion between two electrons occupying the same spatial orbital. The dissociation limit, where the wavefunction is described by \eqref{heitler} can be thought of as non-perturbative, since the Coulomb interaction is much larger than the hybridization, which gives rise to the delocalized $\sigma$-orbitals. The HF and KS-DFT results are far from the desired accuracy in such approaches and it has proven a highly non-trivial task to construct exchange-correlation functionals, that can describe the strong static correlation in this system.

One exception is the Random Phase Approximation (RPA), which was demonstrated to produce the correct dissociation limit of the N$_2$ molecule by Furche.\cite{furche} Subsequently, the RPA has been shown to dissociate several diatomic molecules (including H$_2$) correctly.\cite{ren_review} In these approaches, the calculations are performed non-self-consistently and the method is then equivalent to perturbation theory to infinite order using a subset of terms in the perturbative expansion. At intermediate distances, however, RPA shows qualitative deviation from the exact dissociation energy curve. In particular, a spurious maximum appears and the RPA energy decays too slowly towards the dissociation limit, whereas the exact result rises monotonically to the dissociation limit as the distance is increased. Attempts to improve upon this within time-dependent DFT\cite{hesselmann1, hesselmann2, olsen_ralda1, olsen_ralda2} (TDDFT), have only resulted in an improvement in the absolute correlation energies, but not eliminated the spurious maximum. On the other hand, from a perturbative point of view it is natural to augment RPA with antisymmetrized terms at each order in the perturbation expansion, which eliminates self-interaction terms in RPA. This correction is referred to as Second Order Screened Exchange (SOSEX) \cite{gruneis} and has been shown to completely deteriorate the good description of static correlation within RPA.\cite{henderson, ren_review, eshuis} Recently, it has been demonstrated that total energies obtained from the GW approximation, cannot dissociate the Hydrogen molecule correctly.\cite{caruso} This is highly surprising, since the perturbative expansions involved in RPA and GW are topologically identical and illustrates the subtle nature of the approximations, which are able to capture the static correlation correctly.

Even for such a simple system, first principles calculations can quickly become rather involved. In order to better understand the physical content of the different approximations, we will therefore employ a Hubbard Hamiltonian as a model for a dimer system with on-site Coulomb interactions. This model was previously applied to analyze the excitation spectrum within RPA, Time-Dependent HF, and TDDFT\cite{ferdi} and it was shown that these approximations provide a poor description of the spectrum in the non-perturbative limit. The Hubbard dimer has also been applied as a toy model to examine models for the electronic self-energy beyond the GW approximation\cite{romaniello1,romaniello2} and it was shown that the GW approximation fails to describe the correlated electronic structure in the non-perturbative limit. The failure of the GW approximation to describe quasi-particle excitations in general Hubbard models with large static correlation has also been demonstrated recently.\cite{kaasbjerg} In Ref. \onlinecite{rebolini}, the performance of the Bethe-Salpeter equation was investigated for a hydrogen molecule and was found to yield unphysical (imaginary) excitation spectrum in the dissociation limit. Similar problems have been shown to occur within TDDFT with semilocal adiabatic exchange-correlation kernels.\cite{gritsenko}

In this paper we perform \textit{ab initio} calculations of the molecular dissociation curve using RPA, TDHF, and BSE. It is shown that BSE performs significantly better than RPA and TDHF despite the fact that the excitation spectrum breaks down in the dissociation limit.
We analyze the Hubbard Hamiltonian for a dimer and show how correlation energies are obtained within the framework of the adiabatic connection and fluctuation-dissipation theorem in the model. We also state the exact eigenstates and ground state energy for the model. Various approximations to the correlation energy within the model is then investigated and compared with \textit{ab initio} results. We start by calculating the exact response function and verify that it yields the correct ground state energy when the correlation energy is evaluated within the adiabatic connection fluctuation-dissipation theorem. We then proceed by examining the RPA, which yields a correct dissociation in the strict atomic limit and show that the SOSEX correction deteriorates this result as expected from first principles calculations.\cite{henderson} Hedin's equations are then used to obtain approximations for the response function beyond RPA and we calculate correlation energies within the TDHF, BSE and TDGW approximations. All these approximations yield the correct dissociation in the strict atomic limit, but only BSE and TDGW are able to produce a monotonous dissociation curve in agreement with the exact results.

The paper is organized as follows. In Sec. \ref{hydrogen} we perform the first principles calculations of the hydrogen dissociation curve within RPA, BSE and TDHF. In Sec. \ref{theory} we introduce the Hubbard Hamiltonian and state the framework, which is used to obtain total energies in terms of the response function. We then proceed to investigate how different approximations for the response function translate into dissociation curves for the dimer.

\section{Dissociation of the Hydrogen molecule}\label{hydrogen}
It is well known that RPA is capable of dissociating certain diatomic molecules correctly.\cite{furche, ren_review} However, at intermediate distances the dissociation curves usually display a spurious maximum, which can differ by more than 1.0 eV from the exact dissociation curve. This has been reported both for the dissociation of H$_2$\cite{ren_review, olsen_rpa2, hesselmann2, henderson} as well as the dissociation of N$_2$.\cite{furche, ren_review} Time-dependent density functional theory calculations with an exact exchange kernel has also been shown to yield such a maximum, but reproduces the correct result in the strict atomic limit. In contrast, the HF and SOSEX corrected RPA calculations overestimate the energy in the dissociation limit by $\sim7$ eV and $\sim3$ eV respectively.\cite{olsen_rpa2, ren_review, henderson}

In the context of DFT, we can define total energies as the sum of the exchange energy (non-interacting plus Hartree-Fock) and a correlation energy, which can be obtained from the adiabatic connection fluctuation dissipation theorem (ACDFT). The expression for the correlation energy is
\begin{equation}\label{Ec_ab_initio}
 E_c=-\frac{1}{2}\int^\infty_{-\infty}\frac{d\omega}{2\pi}\int_0^1d\lambda\text{Tr}[v\chi^\lambda(\omega)-v\chi^0(\omega)],
\end{equation}
where $\chi^\lambda$ is the interacting response function at coupling strength $\lambda$, which need to be approximated from either TDDFT \cite{olsen_ralda1, olsen_ralda2} or many-body perturbation theory. In this work we will turn to many-body perturbation theory and Hedin's equations. In the present section we will just show the results of the \textit{ab initio} calculations and in Sec. \ref{theory} we will go through the theory leading to the different approximations for $\chi$ in more detail.

The most famous approximation for $\chi$ is the RPA where one assumes a non-interacting polarization function. The theory and implementation has already been discussed in detail \cite{ren_review, olsen_rpa2} and the RPA dissociation curve of H$_2$ is well known. However, very limited work have been dedicated to going beyond RPA in the context of total energies and many-body perturbation theory. In contrast, for optical excitations in semiconductors, the Bethe-Salpeter equation (BSE) provides a natural "beyond RPA" method that incorporated electron-hole interactions. The BSE is implemented in several existing electronic structure codes, but it is usually applied to analyze the influence of electron-hole interactions on excited states and will typically just return the imaginary part of the macroscopic dielectric function, which is calculated from the eigenvalues and eigenstates of an excitonic Hamiltonian. It is, however straightforward to calculate the full response function from the eigenstates and eigenvalues and the BSE correlation energy can then be obtained by performing a coupling constant integration. We have implemented such a scheme in the electronic structure code GPAW \cite{gpaw-paper}, which already has a fully functioning response part that allows one to calculate excited state properties within the BSE approximation \cite{jun_response, jun_bse}. We apply the usual static RPA approximation for $W$, but not the Tamm-Dancoff approximation. The full time-ordered response function is then constructed in a plane wave basis as
\begin{equation}\label{bse_response}
 \chi^\lambda_{\mathbf{G}\mathbf{G}'}(\omega)=\frac{1}{\Omega_{BZ}}\sum_{S,S'}n_{\mathbf{G}S}\chi^\lambda_{SS'}(\omega)n^*_{\mathbf{G}'S'}
\end{equation}
where
\begin{align}
 \chi^\lambda_{SS'}(\omega)&=f_{S'}\sum_{\alpha,\alpha'}\frac{A^\lambda_{S\alpha}[N^\lambda_{\alpha,\alpha'}]^{-1}A^{\lambda*}_{S'\alpha'}}{\omega-E_\alpha^\lambda}\label{chi_omega}\\
N^\lambda_{\alpha,\alpha'}&=\sum_SA^{\lambda*}_{S\alpha}A^\lambda_{S\alpha'}
\end{align}
and $E_\alpha^\lambda$ and $A_\alpha^\lambda(S)$ are eigenvalues and eigenstates of the BSE Hamiltonian at coupling strength $\lambda$. Here $S$ represents an electron-hole excitation $\psi_S(\mathbf{r}_h,\mathbf{r}_e')=\psi_m(\mathbf{r}_h)\psi_n(\mathbf{r}_e)$, $f_S=f_n-f_m$ is the "excitation occupation", and $n_{\mathbf{G}S}$ is the plane wave representation of the pair density $\psi_S(\mathbf{r},\mathbf{r})$.

For the time-ordered response function, we subtract or add a infinitesimal imaginary part from the eigenvalues in Eq. \eqref{chi_omega} depending on the sign of the real part. From the structure of the BSE Hamiltonian it is straightforward to see that if $E$ is an eigenvalue with eigenvector $(\mathbf{v}_1, \mathbf{v_2})$, where $\mathbf{v}_1(\mathbf{v}_2)$ represents electron-hole(hole-electron) transitions, then $-E^*$ is also an eigenvalue with eigenvector $(\mathbf{v}_2^*, \mathbf{v}_1^*)$. It then follows that the response function will decay as $\omega^{-2}$ for $\omega\rightarrow\infty$. An explicit expression for the correlation energy \eqref{Ec_ab_inition} can then be obtained by closing the contour in the lower half plane and using that the poles of the BSE response function are in the lower half plane when the real part of the $E_S$ is positive. Noting that the non-interacting response function can be written in the form of Eq. \eqref{bse_response} with $\chi_{SS'}=f_S\delta_{SS'}/(\omega-E_S)$ we obtain
\begin{align}\label{E_c_ab}
&E_c^{BSE}=\frac{-1}{2\Omega_{BZ}}\sum_{\mathbf{G},S,S'}f_{S'}n_{\mathbf{G}S}n^*_{\mathbf{G}S'}v_\mathbf{G}\\
&\times\Big[\delta_{SS'}\theta(E_S)-\int_0^1d\lambda\sum_{\alpha,\alpha'}A^\lambda_{S\alpha}[N^\lambda_{\alpha,\alpha'}]^{-1}A^{\lambda*}_{S\alpha'}\theta(\text{Re}E_\alpha^\lambda)\Big]\notag.
\end{align}
Note that this expression allows us to calculate RPA and TDHF correlation energies as well as BSE correlation energies depending on which kernel is used to set up the Hamiltonian.

For our \textit{ab initio} calculations of the hydrogen dissociation curve we have used plane wave cutoffs of 100 eV and 150 eV and extrapolated the results to infinite cutoff assuming that $E_c(E_{cut})\sim E_c(\infty)+AE_{cut}^{-3/2}$. This is not a very accurate procedure and we cannot claim that our BSE calculations are completely converged. However, the static correlation associated with molecular dissociation involves rather large energies and for our purpose, converging the calculations to within 0.2 eV is sufficient. For large supercells or large cutoff energies, it becomes a computationally demanding task to set up and diagonalize the BSE Hamiltonian and with the present implementation, it was not possible to converge the calculations to more than 0.2 eV. In all calculations we have used non-interacting orbitals and eigenenergies obtained with LDA and set the number of states in the initial Kohn-Sham calculation equal to the number of plane waves defined by the cutoff. The coupling constant integration was performed using 8 Gauss-Legendre points. To assert the validity of the implementation and the convergence parameters, we have compared RPA calculations using the expression \eqref{E_c_ab} with a standard, well documented RPA implementation in GPAW.\cite{olsen_rpa1, olsen_rpa2, jun_rpa} This method is based on a direct solution of the two-point Dyson equation for the RPA response function and is performed with an analytical coupling constant integration and numerical frequency integration along the imaginary axis. It was verified that RPA energies obtained with the two methods are identical.
\begin{figure}[tb]
	\includegraphics[width=7.5 cm]{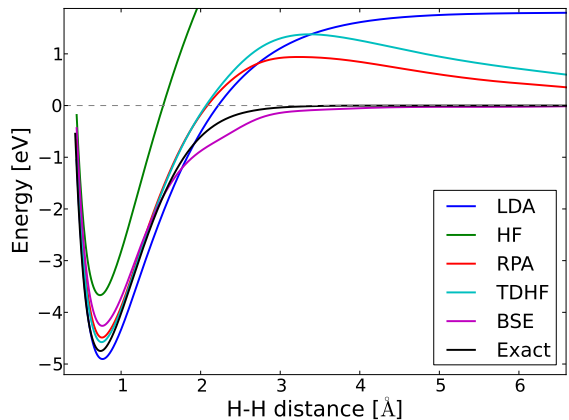}
\caption{(color online). Dissociation curves of H$_2$ calculated from first principles within various approximations. The energy curves have been subtraced the energy of two isolated H atoms within each of the approximations.}
\label{fig:H2}
\end{figure}

The results of the simulations are shown in Fig. \ref{fig:H2}. The LDA and HF energy curves overshoot dramatically in the dissociation limit whereas the RPA, TDHF and BSE curves reproduce the correct dissociation. In addition, the first principles BSE energy curve does not display a spurious maximum and converges rapidly and monotonically towards the dissociation limit. To our knowledge, no other approximation from many-body perturbation theory has been able to reproduce this dissociation curve. It has previously been shown that the excitation energies of H$_2$ can become imaginary in the dissociation limit \cite{rebolini}, which seems to indicate that the theoretical description breaks down. However, the correlation energy is always real and in the context of total energies, the appearance of complex eigenvalues just means that the associated state will not contribute to the correlation energy.

\begin{figure}[tb]
	\includegraphics[width=7.5 cm]{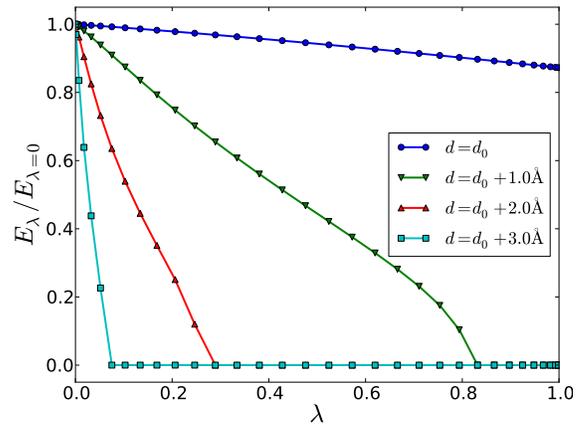}
\caption{(color online). $\lambda$-dependence of the BSE correlation energy at different interatomic distances ($d_0=0.7$ {\AA}). We only consider the interacting part of the correlation energy defined as $E_\lambda=\int d\omega\text{Tr}[v\chi^\lambda(\omega)]$. $E_\lambda$ vanishes when the poles of $\chi^\lambda$ becomes imaginary.}
\label{fig:E_lambda}
\end{figure}
To see this more clearly (and facilitate comparison with the Hubbard model below), we have obtained the eigenvalues of BSE calculations with a single unoccupied band as a function of interatomic distance. At equilibrium distance ($d$=0.7 {\AA}), the two eigenvalues are real and situated at $\pm 7$ eV. When the distance between the atoms is increased, the eigenvalues approach zero, which is reached at d=1.0 {\AA}. After this point the two eigenvalues become a purely imaginary conjugated pair and the absolute value increases steadily as the interatomic distance is increased. The integral of the interacting response function vanishes as soon as the BSE Hamiltonian acquire imaginary poles. This is shown in Fig. \ref{fig:E_lambda} where we have plotted $E_\lambda=\int d\omega\text{Tr}[v\chi^\lambda]$ for different values of the interatomic distance. At $d$=1.0 {\AA} the poles of the interacting response function becomes imaginary and the frequency integral vanishes. However, at larger distances, the interacting response function still contributes to the correlation energy due to the coupling constant integration. If we compare this figure with a corresponding one for RPA calculations \cite{hesselmann1}, we see that the the spurious maximum of RPA can be attributed to the fact that $E^{RPA}_\lambda$ vanishes too slowly for large interatomic distances. Although the kinks in Fig. \ref{fig:E_lambda} may look a bit unphysical, they are in fact responsible for the accurate description of the dissociation in the BSE approximation. We will discuss correlation energies within the BSE approximation further in the context of the Hubbard model below.

Before delving into the Hubbard model for a correlated dimer, we will pause to compare our results to previous calculations of the Hydrogen molecule in the dissociation limit. The RPA potential energy curve has been reported several times in the litterature \cite{ren_review, olsen_rpa2, hesselmann1, henderson} and the results agree very well with our simulations. They all reproduce the exact atomic limit while showing a spurious maximum at intermediate separations. Our TDHF calculations also agree very well with previous calculations \cite{hesselman1} showing a pronounced maximum at intermediate distances. Unfortunately, our implementation does not allow for SOSEX calculations, but we will show below that the Hubbard model yields qualitative agreement with previous calculations \cite{ren_review, henderson}. The BSE excitation spectrum has been reported in Ref. \onlinecite{rebolini} and the appearance of imaginary poles at intermediate separation is in good agreement with the present results, although the BSE calculations in Ref. \onlinecite{rebolini} were performed on top of Hartree-Fock orbitals, whereas the present results are based on LDA.

We should also mention that the correlation energy can also be obtained from the interacting Green functions using the Migdal-Galitski formula. To this end, the GW approximation appears to contain the exact same physics as RPA. However, subtle differences make the two approaches deviate and the GW approach does not reproduce the exact atomic limit correctly \cite{caruso}. Furthermore, going beyond the RPA approximation the correspondence between the Green function method and ACFDT is no longer completely clear, since the self-energy $\Sigma=GW\Gamma$ contains explicit vertex corrections in addition to the vertex-corrected response function entering through $W$.
In fact, it was shown in Ref. \onlinecite{romaniello1} that the exlicit vertex corrections are the most crucial in order to obtain accurate excitation spectra.

\section{Hubbard model}\label{theory}
To elucidate the physical contents of the dissociation curves obtained from the various approximations above, we now turn to the Hubbard model for a dimer. The model will also allow us to unravel the problems associated with the SOSEX correction to RPA and explore the time-dependent GW approximation, which goes beyond the BSE approximation. The Hubbard dimer model is defined by the Hamiltonian
\begin{align}\label{Hamiltonian}
H=\varepsilon_0\sum_{\sigma,i}c_{i\sigma}^\dag c_{i\sigma}-&t\sum_{i\neq j}\sum_\sigma c_{i\sigma}^\dag c_{j\sigma}\\
+ &\frac{U}{2}\sum_{i}\sum_{\sigma\sigma'}c_{i\sigma}^\dag c_{i\sigma'}^\dag c_{i\sigma'} c_{i\sigma}.\notag
\end{align}
The parameter $t$ represents hopping matrix elements between neighboring sites and $U$ is the Coulomb repulsion between electrons occupying the same site. Two distinct limits will be of interest in the following. First, the perturbative limit where $U\ll t$ and $U$ can be treated as a perturbation to the non-interacting system. Second, the atomic (or non-perturbative) limit where $U\gg t$ and the exact eigenstate is ill-described by the non-interacting Slater determinant (static correlation). In the present work, we will only focus on a dimer (molecule) where $\{i,j\}$ runs over two sites. We can then regard $t$ as a measure of inverse bond length and the dissociation (atomic) limit will correspond to $t\rightarrow0$. On the other hand, at typical equilibrium distances, $t$ is comparable to $U$ and Coulomb interactions can often be included perturbatively.

In appendix \ref{app_hubbard} we give a brief summary of the exact eigenstates of the model \eqref{Hamiltonian}. We also summarize how correlation energies can be derived from the density response function within the adiabatic-connection and fluctuation-dissipation theorem. To obtain approximations for the response function we write it in terms of the irreducible response $P_{ij,\sigma\sigma'}(\omega)$:
\begin{align}\label{dyson}
 \chi(i\omega)=P(i\omega)+P(i\omega)v\chi(i\omega),
\end{align}
where we suppressed spin and site indices. This equation for $\chi$ and $P$ follows from the definitions of these quantities as the density response to an external potential and the density response to the total (external plus Hartree) potential. Using that the Coulomb interaction is independent of spin, it follows that the spin-summed quantities satisfy
\begin{align}
 \widetilde\chi(i\omega)=\widetilde P(i\omega)+U\widetilde P(i\omega)\widetilde\chi(i\omega),
\end{align}
or
\begin{align}
 \widetilde\chi(i\omega)=[1-U\widetilde P(i\omega)]^{-1}\widetilde P(i\omega).
\end{align}
Various approximations for the irreducible response can be derived in the context of Hedin's equation which are summarized in appendix \ref{app_hedin}.

\subsection{Non-interacting response function}\label{noninteracting}
We start by evaluating the non-interacting polarizability function $P^0(12)=-iG^0(12)G^0(12)$. The Fourier transform is then given by
\begin{align}
P^0_{ij,\sigma\sigma'}(\omega)=i\int\frac{d\omega'}{2\pi}G^0_{ij,\sigma\sigma'}(\omega')G^0_{ji,\sigma\sigma'}(\omega+\omega').
\end{align}
The non-interacting Green function can be evaluated from its spectral representation after having diagonalized the $N=1,3$ sectors of the non-interacting Hamiltonian. The result is
\begin{align}
G^0_{ij,\sigma\sigma'}(\omega)=\frac{\delta_{\sigma\sigma'}}{2}\Big[\frac{(-1)^{i-j}}{\omega-(\varepsilon_0+t)+i\eta}+\frac{1}{\omega-(\varepsilon_0-t)-i\eta}\Big],
\end{align}
from which we obtain the non-interacting polarizability:
\begin{align}\label{P_0_real}
P^0_{ij,\sigma\sigma'}(\omega)=\frac{\delta_{\sigma\sigma'}(-1)^{i-j}}{4}\Big[\frac{1}{\omega-2t+2i\eta}-\frac{1}{\omega+2t-2i\eta}\Big].
\end{align}
In the following it will often be convenient to work with imaginary frequencies and we obtain the polarization function by analytic continuation:
\begin{align}\label{P_0}
P^0_{ij,\sigma\sigma'}(i\omega)=-(-1)^{i-j}\delta_{\sigma\sigma'}\frac{t}{\omega^2+4t^2}.
\end{align}

\subsection{Exact response function}
For later reference it will also be useful to calculate the exact response function. This is most easily done by expressing the density-density correlation function in its spectral represention and use the eigenstates Eq. \eqref{eigen_states}. The result is
\begin{align}\label{exact_response}
\chi_{ij,\sigma\sigma'}(i\omega)=-(-1)^{i-j}\Big[&\frac{\frac{U+c}{2a^2}}{\omega^2+(U+c)^2/4}\\
+&\frac{(-1)^{1-\delta_{\sigma\sigma'}}\frac{8t^2}{a^2(c-U)}}{\omega^2+(U-c)^2/4}\Big]\notag,
\end{align}
where $a$ and $c$ were defined in Eq. \eqref{abc}. The second term originates from the triplet state $|\psi_3\rangle$ and does not contribute to the correlation energy which becomes:
\begin{align}
E_c&=-U\int_0^1d\lambda\int_0^\infty\frac{d\omega}{2\pi}\sum_i\Big[\widetilde\chi_{ii}^\lambda(i\omega)-\widetilde\chi_{ii}^0(i\omega)\Big]\notag\\
&=U\int_0^1d\lambda\Big[\frac{2}{a_\lambda^2}-\frac{1}{2}\Big]=2t-\frac{c}{2}.
\end{align}
This is identical to the result obtained from exact diagonalization (Eqs. \eqref{eigen_values} and \eqref{E_c_def}) as it should be. Below we will calculate the interacting response function and correlation energies within various approximations. Note, that the correlation energy always acquires a finite contribution from the non-interacting response function, which is composed of an electronic transition between a delocalized doubly occupied $\sigma$-orbital ($|\psi_0^{\lambda=0}\rangle$) and the $|\psi_4\rangle$ state. In the atomic limit, finite order perturbation theory can never produce a term that exactly cancels the $U/2$ originating from first order perturbation theory.

It is interesting to note that in this framework, the ground state energy can be decomposed into contributions originating from transitions to excited many-body eigenstates integrated along the adiabatic connection. The derivation of the response function \eqref{exact_response} shows that only the transition from $|\psi_0^\lambda\rangle$ to $|\psi_4\rangle$ where both electrons are always located on the same site, contributes to the ground state energy. In the atomic limit the electrons are never localized on the same site in the ground state and the transition can be regarded as a pure charge transfer excitation. This implies that the transition matrix elements of the density operator should vanish in the atomic limit, which can be verified from the numerator of \eqref{exact_response} after having decomposed the expression into two terms with simple poles. In general, any good approximation of an interacting response function, should have the property that the amplitude of the charge transfer excitation vanishes in the atomic limit. This will show up as a vanishing numerator in the response function and translate into a correct correlation energy in the atomic limit. In contrast, the numerator of the non-interacting response function \eqref{P_0_real} does not vanish since the non-interacting ground state always gives a probability of $1/2$ for finding both electrons at the same site.

From the response function, we also calculate the exact polarizability which becomes
\begin{align}\label{exact_P}
P_{ij,\sigma\sigma'}(i\omega)=&-\frac{(-1)^{i-j}\frac{U+c}{2a^2}}{\omega^2+(U+c)^2/4-4U(U+c)/a^2}\notag\\
&-\frac{(-1)^{i-j}(-1)^{1-\delta_{\sigma\sigma'}}\frac{8t^2}{a^2(c-U)}}{\omega^2+(U-c)^2/4}.
\end{align}
The most important thing to note, is the fact that it is non-diagonal in spin. As we will see later, a rather advanced approximation (TDGW) is needed in order to introduce non-vanishing off-diagonal spin elements in the polarization function. It is also interesting to note that only the singlet pole corresponding to a transition to $|\psi_4\rangle$, becomes renormalized with respect to the poles of the response function. This is because the Coulomb interaction is independent of spin and therefore all term that have a sign change associated with a spin flip is eliminated from the Dyson equation \eqref{dyson}.

\subsection{RPA}
The Random Phase Approximation (RPA) is obtained by taking $P(12)=-iG^0(12)G^0(21)$. Using Eqs. \eqref{dyson} and \eqref{P_0}. The response function is
\begin{align}\label{rpa_response}
\chi_{ij,\sigma\sigma'}^{RPA}(i\omega)&=\chi^0_{ij,\sigma\sigma'}(i\omega)+\frac{2Ut^2(-1)^{i-j}}{(\omega^2+4t^2+4Ut)(\omega^2+4t^2)}\notag\\
&=-\frac{t(-1)^{i-j}}{2}\Big[\frac{(-1)^{1-\delta_{\sigma\sigma'}}}{\omega^2+4t^2}+\frac{1}{\omega^2+4t^2+4tU}\Big]
\end{align}
and the correlation energy becomes
\begin{align}
E_c^{RPA}&=-U\int_0^1d\lambda\int_0^\infty\frac{d\omega}{2\pi}\frac{16\lambda Ut^2}{(\omega^2+4t^2+4\lambda Ut)(\omega^2+4t^2)}\notag\\
&=-\frac{U}{2}\int_0^1d\lambda\Big[1-\frac{1}{\sqrt{1+\lambda U/t}}\Big]\notag\\
&=-\frac{U}{2}-t(1-\sqrt{1+U/t}).
\end{align}
Note that for small $U$, the leading term in this expression becomes $\sim -U^2/8t$, which is only half the exact second order contribution Eq. \eqref{E_U}. This is due to the lack of second order exchange in the RPA approximation. On the other hand, for $t\rightarrow0$ the expression nicely cancels the $U/2$ contribution from first order perturbation theory and total energy reduces to the non-interacting one in exact atomization limit.

The RPA response function \eqref{rpa_response} has a structure very similar to the exact response function \eqref{exact_response}. It consists of two terms representing transitions to $|\psi_3\rangle$ and $|\psi_4\rangle$, which are a triplet and singlet respectively. The triplet excitation energy does not become renormalized with respect to the non-interacting transition energy, but does not contribute to the correlation energy. The singlet excitation contains all the correlation energy and its frequency integral vanishes for $t\rightarrow0$, which ensures the correct dissociation limit. It is instructive to represent the RPA response function in terms of electron-hole transitions. In the atomic limit the eigenstates of the electron-hole Hamiltonian becomes $|eh\rangle=(|1\rangle_e\otimes|2\rangle_h-|2\rangle_e\otimes|1\rangle_h)/\sqrt{2}$, where $|i\rangle$ is a single-particle state at site $i$ (see appendix \ref{app_hamilton}). This indicates that the electronic transition contributing to the correlation energy becomes a pure charge transfer excitation in the atomic limit and the interacting response function thus vanishes.

Although the RPA can reproduce the exact atomic limit, the asymptotic behavior is very different from the exact result. The total RPA ground state energy approaches zero as $\sim\sqrt{Ut}$, whereas the exact result vanishes much faster as $-4t^2/U$ (see Eq. \eqref{E_t}). In fact, the exact ground state energy approaches zero monotonically with decreasing $t$, whereas the RPA ground state energy has a maximum at $t_M=4(1-\sqrt{5/6})^2U\approx0.03U$.

The $\sim\sqrt{t}$ scaling of the correlation energy in the atomic limit can already be recognized from the response function \eqref{rpa_response}. Decomposing the second term into two terms with single poles, we see that the numerators become proportional to $\sim\sqrt{t}$, which indicates that the RPA transition matrix elements scales as $t^{1/4}$.

\subsection{SOSEX}
The SOSEX correction to RPA is inspired by the expansion of the RPA energy in terms of Feynman diagrams. The RPA can then be written as an infinite sum of direct ring diagrams which can be re-summed to yield a screened direct second order term. It is natural to add the associated exchange terms at each order in the perturbative expansion. In particular, this will ensure that the correlation energy of a single electron vanishes. Such a contribution is called second order screened exchange (SOSEX) and the correction to the RPA correlation energy is
\begin{widetext}
\begin{align}
\Delta E_c^{SOSEX}&=-\frac{U}{2}\int_0^1d\lambda\int\frac{d\omega}{2\pi}\text{Im}\sum_{i}\widetilde P_{ii}^{1,\lambda}(\omega)
\end{align}
where
\begin{align}\label{P1}
P^1_{ij,\sigma\sigma'}(\omega)=\delta_{\sigma\sigma'}\sum_{kl}\int\frac{d\omega_1d\omega_2}{(2\pi)^2}G^0_{jk}(\omega_1)G^0_{lj}(\omega+\omega_1)W_{kl}(\omega_1-\omega_2)G^0_{il}(\omega+\omega_2)G^0_{ki}(\omega_2),
\end{align}
and the screened interaction is
\begin{align}
W_{kl}(\omega)=U\delta_{ij}+\frac{(-1)^{i-j}U^2t}{\omega^2-h^2},\qquad h=\sqrt{4t^2+4Ut}-i\eta\frac{4t+U}{\sqrt{4t^2+2Ut}}.
\end{align}
Here we have also written $G^0_{ij}=G^0_{ij,\uparrow\uparrow}=G^0_{ij,\downarrow\downarrow}$ for short. For later reference we perform the integrations, which yield the expression
\begin{align}\label{P1_int}
P^1_{ij}(i\omega)=-\frac{2Ut^2(-1)^{i-j}}{(\omega^2+4t^2)^2}+\frac{U^2t}{h(2t+h)}\bigg[\frac{(-1)^{i-j}}{\omega^2+4t^2}+\frac{1}{\omega^2+(2t+h)^2}\bigg],
\end{align}
where $P^1_{ij,\sigma\sigma'}=P^1_{ij}\delta_{\sigma\sigma'}$. However, the SOSEX correction to the correlation energy is obtained much easier by comparing with the RPA correlation energy, which can be written
\begin{align}
E_c^{RPA}=2U\text{Im}\sum_{ikl}\int_0^1d\lambda\int\frac{d\omega d\omega_1d\omega_2}{(2\pi)^3}G^0_{ik}(\omega_1)G^0_{ki}(\omega_1-\omega)W^\lambda_{kl}(\omega)G^0_{jl}(\omega_2-\omega)G^0_{lj}(\omega_2)\notag.
\end{align}
Here a factor of four originates from the spin summations. Rewriting the SOSEX correction a bit gives
\begin{align}
\Delta E_c^{SOSEX}&=-\frac{U}{2}\text{Im}\sum_{i}\int_0^1d\lambda\int\frac{d\omega}{2\pi}\widetilde P_{ii}^{1,\lambda}(\omega)\notag\\
&=-U\text{Im}\sum_{ikl}\int_0^1d\lambda\int\frac{d\omega d\omega_1d\omega_2}{(2\pi)^3}G^0_{ik}(\omega_1)G^0_{li}(\omega+\omega_1)W^\lambda_{kl}(\omega_1-\omega_2)G^0_{il}(\omega+\omega_2)G^0_{ki}(\omega_2)\notag\\
&=-U\text{Im}\sum_{ikl}\int_0^1d\lambda\int\frac{d\omega d\omega_1d\omega_2}{(2\pi)^3}G^0_{ik}(\omega_1)G^0_{ki}(\omega_1-\omega)W^\lambda_{kl}(\omega)G^0_{il}(\omega_2-\omega_2)G^0_{li}(\omega_2)\notag
\end{align}
\end{widetext}
where the last equality was obtained by the substitutions $\omega_2\rightarrow\omega_2-\omega$, $\omega\rightarrow\omega+\omega_2-\omega_1$, and $\omega_2\rightarrow\omega_2-\omega$. Comparing with the RPA expression then yields
\begin{align}
\Delta E_c^{SOSEX}=-\frac{E_c^{RPA}}{2}.
\end{align}
It is straightforward to verify that this is also obtained by explicit integration of Eq. \eqref{P1_int}. The factor of two originates from the fact that the SOSEX contribution is diagonal in spin whereas the RPA contribution is independent of spin. The total correlation energy is then
\begin{align}
E_c^{RPA+SOSEX}&=E_c^{RPA}/2\\
&=-\frac{U}{4}-\frac{t}{2}(1-\sqrt{1+U/t})\\
&=-\frac{U^2}{16t}+\frac{U^3}{32t^2}+\ldots,\qquad t\le U.
\end{align}
The SOSEX corrected correlation energy is exact within second order perturbation theory as expected from its construction. However, the expression does not cancel the first order contribution of $U/2$ in the non-perturbative limit and therefore predicts a wrong dissociation energy of $E^{RPA+SOSEX}\rightarrow2\varepsilon0+U/4$ for $t\rightarrow0$. This is most likely due to the fact that the SOSEX corrected correlation energy cannot be written as an integral over a \textit{reducible} response function. As will be shown below, the SOSEX correction corresponds to an \textit{irreducible} polarization function and therefore fits oddly into the present scheme of calculating correlation energies from the adiabatic connection.

\subsection{1W correction}
Whereas the SOSEX correction seems like a natural extension of RPA from the point of view of many-body perturbation theory, we will now try to go beyond the RPA starting from Hedin's equations Eq. \eqref{G}-\eqref{Gamma}. The RPA for the polarizability can be obtained by taking $\delta\Sigma/\delta G=0$. Instead we will now take
\begin{align}
 \frac{\delta\Sigma(12)}{\delta G(34)}=iW(12)\delta(13)\delta(24),
\end{align}
which follows from the RPA self-energy $\Sigma=G^0W^0$ if one neglects that $W$ depends on $G^0$. Iterating $\Gamma$ one time then yields
\begin{align}
 P^{1W}(12)=P^0(12)+P^1(12),
\end{align}
where
\begin{align}\label{P1_again}
 P^{1}(12)=\int d34G(13)G(41)W(34)G(32)G(24)
\end{align}
We recognize that this is exactly the SOSEX function appearing in Eq. \eqref{P1}, however, since this is now part of an irreducible polarizability, we should evaluate the correlation energy from the Dyson equation \eqref{dyson} with $P=P^0+P^1$. The correlation energy then becomes (suppressing integrations)
\begin{align}\label{SOSEX_correction}
E_c^{1W}=&E_c^{RPA+SOSEX}\\
&+2\text{Tr}[vP^0vP^1]+3\text{Tr}[vP^0vP^0vP^1]+\ldots\notag.
\end{align}
Here the RPA energy is the sum of all terms not containing $P^1$ and the SOSEX correction is $\text{Tr}[vP^1]$. From this expression it seems inconsistent not to include all the cross terms involving both $P^0$ and $P^1$ as well as the SOSEX correction. Since we have an explicit expression for $P^1$ \eqref{P1_int}, we can insert this into the Dyson equation \eqref{dyson} and obtain the full correlation energy. However, the resulting expression for the reducible response function is rather complicated so we will start by calculating the correlation energy resulting from a single bare Coulomb interaction ($W(\omega)=V$) in the polarization. The spin-summed response with this approximation becomes
\begin{align}\label{dyson_1V}
\widetilde\chi_{ij}^{1V}(i\omega)=\frac{(-1)^{i-j}\widetilde P^{1V}_{ii}(i\omega)}{1-2U\widetilde P^{1V}_{ii}(i\omega)}
\end{align}
with
\begin{align}
\widetilde P^{1V}_{ii}(i\omega)=-\frac{2t}{\omega^2+4t^2}-\frac{4Ut^2}{(\omega^2+4t^2)^2}.
\end{align}
The trace of the response function evaluated at real frequencies can then be written
\begin{align}
\sum_i&\widetilde\chi_{ii}^{1V}(\omega)=-\frac{4t(4t^2+2tU-\omega^2)}{(\omega^2-\omega_+^2)(\omega^2-\omega_-^2)}
\end{align}
with
\begin{align}
\omega^2_\pm=4t^2+(2\pm2i)tU.
\end{align}
We see that the response function acquires complex poles even though the polarization function has real poles. This indicates that the approximation fails dramatically. However, we may still define the correlation energy in terms of the imaginary part of the response function even though this has complex poles corresponding to a non-Hermitian Hamiltonian. The physical interpretation of the poles becomes obscure, but the correlation energy is real and well defined. In this particular case the poles constitute two complex conjugate pairs, which renders the response function real and the correlation energy defined from the dissipation fluctuation theorem vanishes. The exact same situation applies if we replace the bare interaction with a screened interaction and calculate the response function from $P^{1W}=P^0+P^1$, with $P^1$ given by Eq. \eqref{P1_again}. One may naively think that a second order expansion (in $U$) of this correlation energy should reproduce the second order RPA+SOSEX energy since only the RPA+SOSEX energy contributes to a second order expansion of Eq. \eqref{SOSEX_correction}. However, a finite Taylor expansion of Eq. \eqref{dyson_1V} in $U$ is not well defined for all frequencies due to the poles in $P$. We find it puzzling though, that the resummation \eqref{dyson_1V} yields complex poles at all values of $U$. This implies that a simple iteration of the equation \eqref{Gamma} is not sufficient if we want a good approximation for $P$ and the situation seems to be similar to the Dyson equation for $\chi$, where a complete resummation of the Dyson equation \eqref{dyson} is needed in order to obtain a good approximation.

\begin{figure}[tb]
	\includegraphics[width=7.5 cm]{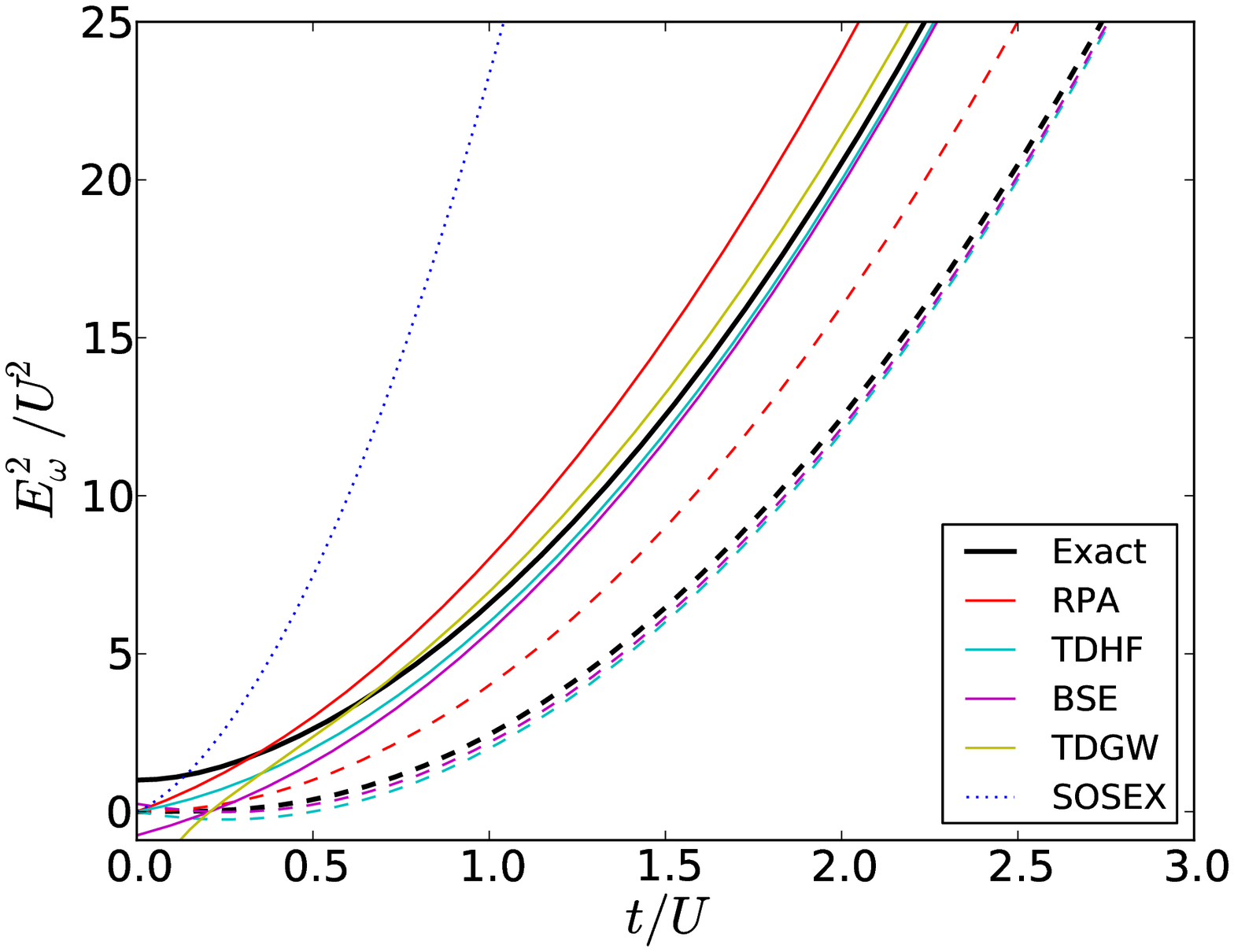}
        \includegraphics[width=7.5 cm]{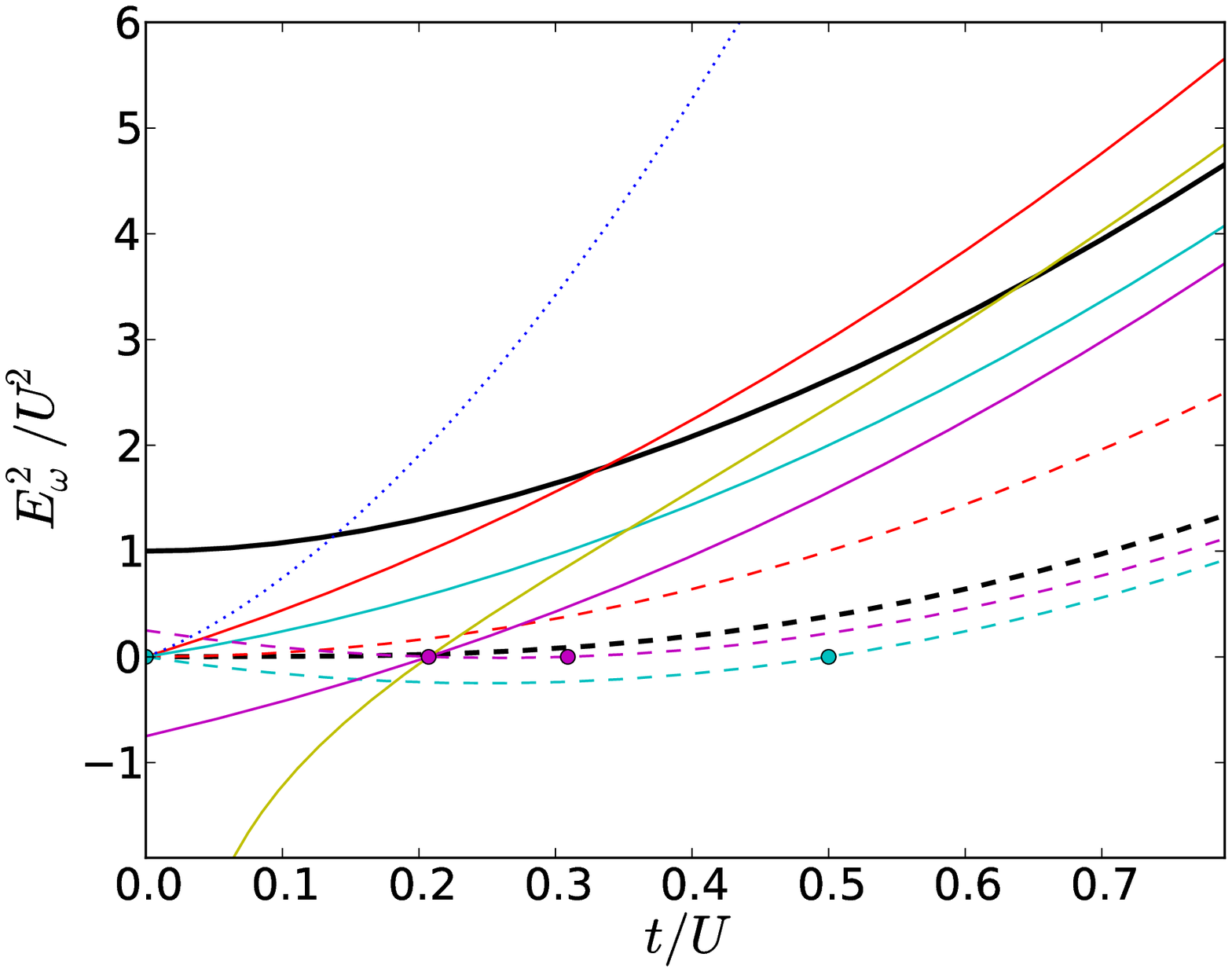}
\caption{(color online). Poles squared of the Hubbard dimer at half filling within different approximation. The bottom figure is a zoom in on the non-perturbative region. The solid lines are singlet excitations and the dashed lines are excitations to the triplet state. The SOSEX pole is marked as a dashed line since it is not a proper response function pole. The non-interacting excitation coincides with the RPA triplet excitation. The zero points of the poles are marked with circles and the negative region between zero points lead to imaginary frequencies.}
\label{fig:poles}
\end{figure}

\subsection{BSE}
Instead of simply including the first order correction in $W$ when calculating $P$, it is possible to explicitly calculate the infinite series of diagrams that generate $P$ when we iterate $\Gamma$ an infinite number of times. In fact, one can show that the polarization function satisfies the four-point Dyson equation
\begin{align}\label{four_dyson}
P(1234)=P^0(1234)+\int d5678P(1256)K(5678)P^0(7834)
\end{align}
where
\begin{align}\label{general_kernel}
P(12)=P(1122),\qquad K(1234)=-i\frac{\delta\Sigma(12)}{\delta G(34)}.
\end{align}
To proceed we will make a static approximation for $K$ and use $\Sigma=G^0W$. Furthermore, if we neglect the $G^0$-dependence in $W$ we obtain the Bethe-Salper approximation for P where
\begin{align}\label{four_kernel}
K^{BSE}(1234)=W(21)\delta(13)\delta(24)\delta(t_1-t_2).
\end{align}
The last delta function comes from the static approximation. It is straightforward to recognize that the BSE polarization is diagonal in spin. The diagonal spin components are can be calculated directly from Eq. \eqref{four_dyson} and yield
\begin{align}
P_{ij}(i\omega)=&-(-1)^{i-j}\frac{t(1-U^2/h^2)}{\omega^2+\omega_1^2}.
\end{align}
and from the two-point Dyson equation we can calculate the BSE response function which becomes
\begin{align}\label{bse_response}
\chi_{ij,\sigma\sigma'}(i\omega)=&\frac{-t(1-U^2/h^2)(-1)^{i-j}}{2}\\
&\times\Big[\frac{1}{\omega^2+\omega_0^2}+\frac{(-1)^{1-\delta_{\sigma\sigma'}}}{\omega^2+\omega_1^2}\Big]\notag
\end{align}
where
\begin{align}
\omega_0^2=&\frac{16t^4 + 40t^3U + 32t^2U^2 + 6tU^3 - 3U^4}{4(t + U)^2},\label{om_bse}\\
\omega_1^2=&\frac{16t^4 + 24t^3U - 6tU^3 + U^4}{4(t + U)^2}.
\end{align}
The poles become purely imaginary in a certain parameter range close to the non-perturbative limit. In particular, $\omega_0$ becomes imaginary for $t/U<(\sqrt{2}-1)/2$ and $\omega_1$ becomes imaginary for $(\sqrt{2}-1)/2<t/U<(\sqrt{5}-1)/4$. It is very interesting that the poles become imaginary exactly at the degeneracy point. This indicates that below this point the lowest singlet and triplet states cross and the response function becomes ill-defined since the reference state is no longer the ground state. However, it should be noted that the poles themselves do not enter the expression for the correlation energy. Rather it is the, matrix elements of the density operator corresponding to the transition associated with a certain pole. It is still interesting though, to compare the transition energies calculated within the different approximations and one would typically expect that the transition matrix elements are well approximated if the poles are accurate. We show the singlet and triplet poles calculated within various approximations in Fig. \ref{fig:poles}. These results are in good agreement with ab initio BSE calculations for the hydrogen molecule.\cite{rebolini}

Again the triplet excitation does not contribute to the correlation energy since it is eliminated in the spin summation. The frequency integral of the response function (evaluated at real frequencies) becomes real for a purely imaginary pole and does not contribute to the correlation energy for $t/U<\lambda(\sqrt{2}-1)/2$. The appearance of imaginary poles signals a breakdown of the BSE response function in the non-perturbative limit. However, from the point of view of the fluctuation dissipation theorem \eqref{fluctuation}, the contribution to the correlation energy from the interacting response function, can be associated with the matrix elements $A=\langle\psi_0|n_i|\psi_4\rangle$. The imaginary poles simple means that $A^{BSE}=0$ for $t/U<(\sqrt{2}-1)/2\approx0.02$, whereas the exact condition is $A\rightarrow0$ for $t/U\rightarrow0$. Due to the coupling constant integration the correlation energy vanishes smoothly and provides an accurate description of the non-perturbative regime (see Fig. \ref{fig:model}).

\subsection{TDHF}
Traditionally, the $W$ function is interpreted as a screened Coulomb potential. However, in molecular systems the interpretation of $W$ is not so clear and should be regarded as a auxiliary function, which should be calculated along with $G$, $P$, $\Sigma$, and $\Gamma$ in order to solve the full many-body problem. For example, the exact $W$ for a one-electron system is $W=v+v\chi_0v$\cite{romaniello1}, which clearly differs from the exact Coulomb interaction although there is no additional electrons to mediate the screening. Since the physical interpretation of $W$ is not completely clear, we may try to simply replace it by the bare Coulomb interaction $V$ in the expression for $\Sigma$. This lead to the Hartree-Fock self-energy $GV$ and the kernel \eqref{four_kernel} becomes equal to (four-point) $V$. The response function derived from this procedure is then called time-dependent Hartree-Fock. The polarization function, which is an infinite series of $V$-ladder diagrams can be re-summed to yield
\begin{align}
P_{ij}(i\omega)=-(-1)^{i-j}\frac{t}{\omega^2+4t^2-2tU},
\end{align}
and the response function becomes
\begin{align}\label{tdhf_response}
\chi_{ij,\sigma\sigma'}(i\omega)=&-\frac{t(-1)^{i-j}}{2}\\
&\times\Big[\frac{(-1)^{1-\delta_{\sigma\sigma'}}}{\omega^2+4t^2-2tU}+\frac{1}{\omega^2+4t^2+2tU}\Big].\notag
\end{align}
Again the structure is very similar to both RPA and BSE. The expression is in fact very similar to the RPA response function \eqref{rpa_response}, the only difference being that the square of the two poles has been shifted by $2Ut$. This means that the TDHF approximation provides a renormalization of both the singlet and triplet excitations and not just the singlet excitation as was the case for RPA. Furthermore the singlet pole is much closer to the exact value than that of RPA as can be seen from Fig. \ref{fig:poles}. In fact, the accuracy of the singlet pole seems to be better than that of BSE, and never becomes imaginary. On the other hand, the triplet pole becomes imaginary for $t<U/2$, and TDHF provides a worse description for this state than BSE. The state does, however, not contribute to the correlation energy, which becomes
\begin{align}
E_c^{TDHF}&=-U\int_0^1d\lambda\int_0^\infty\frac{d\omega}{2\pi}\frac{4t}{\omega^2+4t^2+2\lambda Ut}-\frac{U}{2}\notag\\
&=-\frac{U}{2}\int_0^1d\lambda\frac{1}{\sqrt{1+\lambda U/2t}}-\frac{U}{2}\notag\\
&=-\frac{U}{2}-2t(1-\sqrt{1+U/2t}).
\end{align}
Note that a Taylor expansion in $U$ gives the correct second order term $E_c^{(2)}=-U^2/16t$ as it should since TDHF is exact to second order. This was also the case for the SOSEX corrected RPA, but in contrast to TDHF, that approximation did not reproduce the correct dissociation limit. However, like RPA the asymptotic behavior in the non-perturbative limit is incorrect and the correlation energy goes as $\sim\sqrt{tU/2}$. Thus it vanishes even slower than RPA and as can be seen from Fig. \ref{fig:model} it also has a spurious maximum which is situated even higher than in the case of RPA. This is perhaps surprising since the TDHF poles are much more accurate than the RPA poles. Apparently, the RPA provides a better approximation for the transition matrix elements than TDHF although TDHF gives much more accurate poles.

\subsection{TDGW}
\begin{figure}[tb]
	\includegraphics[width=7.5 cm]{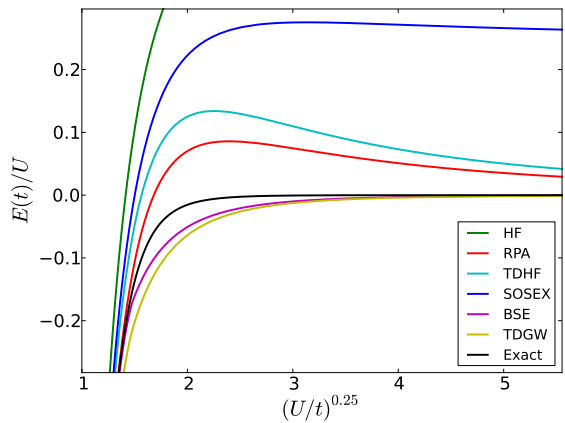}
\caption{(color online). Ground state energy of the Hubbard dimer calculated in different approximations with $\varepsilon_0=0$. Only HF and SOSEX do not give the correct dissociation limit of $E=0$, but yields $E=U/2$ and $E=U/4$ respectively. RPA and TDHF yields large spurious maxima and decays very slowly towards dissociation, whereas BSE and TDGW gives a rapid monotonous rise to dissociation in accordance with the exact results.}
\label{fig:model}
\end{figure}

The BSE polarization was derived by taking the $GW$ approximation for the self-energy and approximating its $G$-functional derivative by $W$. However, using a non-interacting polarization function, $W$ has a rather simple dependence on $G$ and we may carry out the functional derivative of $W$. This gives (integrations suppressed)
\begin{align}
\frac{\delta W}{\delta G}=V\frac{\delta P^0}{\delta G}W+VP^0V\frac{\delta P^0}{\delta G}W+\ldots=W\frac{\delta P^0}{\delta G}W.
\end{align}
The functional derivative of the non-interacting polarization gives two terms and the the final results becomes
\begin{align}\label{tdgw_kernel}
K^{TDGW}(1234)=&W(21)\delta(13)\delta(24)\\
&-iG(12)W(23)G(34)W(41)\notag\\
&-iG(12)W(24)G(43)W(31).\notag
\end{align}
The resulting polarization function is exact within the $GW$ approximation and is referred to as time-dependent $GW$ (TDGW). Again we will make a static approximation for the kernel, in order to solve a single frequency four-point Dyson equation. An intriguing property of this approximation is that it does not yield a spin-diagonal polarization function. We know that the exact polarization is not diagonal in spin, but neither RPA, $P^{1W}$, BSE of TDHF can yield off-diagonal spin blocks for the polarization. The spin summed response function is very similar to the BSE case except that the poles are a bit more complicated. Here we just give the spin summed response function which is
\begin{align}\label{tdgw_response}
\widetilde\chi_{ij}(i\omega)=\frac{-2t(1-U^2/h^2)(-1)^{i-j}}{\omega^2+\omega_0^2}
\end{align}
where
\begin{align}
\omega_0^2=&\big[h^2-U^2\big]\big[4h^2t^2(h+t)^2+2h^2tU(h+t)^2\\
&+2(h+t)^2(h^2+2t^2)U^2-4h^2tU^3\big]/h^4(h+t)^2\notag.
\end{align}
Again the pole becomes imaginary when $t/U<(\sqrt{2}-1)/2$ and the results of the approximation are very similar to BSE. In fact, Figs. \ref{fig:poles} and \ref{fig:model} indicates that the TDGW results are slightly worse than the BSE results. This could be related to the fact that the static approximation in TDGW is somewhat more drastic than the static approximation in BSE. The BSE four-point kernel only depends on a single frequency, whereas the Fourier transform of the TDGW kernel \eqref{tdgw_kernel} depends on three independent frequencies, which are all set to zero in the static approximation.

\section{Conclusion}
We have performed \textit{ab initio} calculations of the dissociation curve of a hydrogen molecule using the RPA and BSE approximations for the response function. Both approximation produce the correct dissociation limit, but the BSE result clearly improve the RPA description at intermediate distances and provides the qualitative correct monotonous rising dissociation curve. For the BSE response function, we see the appearance of complex poles at strong coupling, but the correlation energy is real and only vanishes exactly in the strict atomic limit. These calculations are two orders of magnitude more time consuming than RPA calculations and is most likely to involved to be useful for routine \textit{ab initio} electronic structure simulations at the moment. Nevertheless, the steady increase in computer power may render the method useful in a few years and there is certainly room for optimization of the algorithm used here to obtain the correlation energies.

To obtain more insight into the physics contained in the different dissociation curves, we have examined various approximations to the correlation energy of a Hubbard dimer within many-body perturbation theory. Comparison with the \textit{ab initio} results show that this simple model is able to capture the qualitative features of a first principles treatment of molecular hydrogen in the dissociation limit. In particular, RPA provides the correct dissociation limit but displays a spurious maximum in the dissociation curves, whereas the SOSEX corrected RPA does not yield the correct dissociation limit. Whereas the SOSEX energy cannot be expressed in terms of an irreducible response function the expression itself corresponds to a polarization function resulting from a single iteration of the vertex equation \eqref{Gamma} with the GW self-energy. This inspired us to define a $\chi^{1W}$ obtained by solving the Dyson equation for $\chi$ using the "SOSEX polarization" $P^{1W}$. However, this approach yields a response function with complex poles in all of parameter space and indicates that a simple iterative approach to the vertex equation does not produce good approximations for the polarization function. Instead, one has to solve the equation for the $\Gamma$ to obtain a fully renormalized vertex, which in turn produces good approximations for the polarization. With this method one obtains TDHF, BSE or TDGW depending on the approximation used for the functional derivative of the self-energy and these approximations all yield the correct dissociation limit. In contrast to TDHF, BSE and TDGW gives rise to monotonously increasing dissociation curves in accordance with the exact result. This implies that the screened interaction $W$ is a much better perturbative quantity than the bare interaction $V$. In the case of metals, this is common knowledge and the physical origin of the screening is well understood. For molecules, however, it is less clear why the screened interaction appears and one should regard it as a auxiliary function which replaces $V$ in Hedin's formulation of many-body perturbation theory. On the other hand, the poles of the singlet terms in the BSE and TDGW response functions become imaginary in the non-perturbative limit. Usually, one would regard this as a breakdown of the theory since it indicates that the spectrum is described by a non-Hermitian Hamiltonian, which indicates that time-evolution is not unitary. In the present context of ground state correlation energies, the response function is just an object which allow us to approximate the correlation function $\langle0|\hat n(\mathbf{r})\hat n(\mathbf{r}')|0\rangle_\lambda$, which is real by definition. The appearance of imaginary poles just implies that this correlation function vanishes at large coupling strength $\lambda$ when $U$ is sufficiently large. However, due to the coupling constant integration, the correlation energy only vanishes exactly in the strict dissociation limit.

In the case of the Hydrogen molecule, the exact dissociation curve can be obtained with the configuration interaction method \cite{wolnie} but for more complicated systems such an approach becomes impossible. In solid state physics the Mott insulators\cite{mott} comprise a good example of systems where mean-field approaches typically fail due to strong static correlation. It will be interesting to see if many-body methods such as those investigated here can provide an accurate description of ground state properties in these systems.

\begin{acknowledgments}
The authors acknowledge support from the Danish Council for Independent Research's Sapere Audie Program, grant no 11-1051390. The Center for
Nanostructured Graphene is sponsored by the Danish National Research Foundation.
\end{acknowledgments}

\appendix
\section{Hubbard model}\label{app_hubbard}
\subsection{Exact diagonalization}
Here we will state the exact eigenvalues and and eigenstates of the Hamiltonian \eqref{Hamiltonian}. The Fock space is spanned by the Hilbert spaces corresponding to $N=0,1,2,3,4$ electrons. We will restrict ourselves to the case of two electrons in the following. The Hilbert space is then spanned by the six Slater determinants:
\begin{align}
N=2:\quad|\uparrow\;\downarrow\rangle,\;|\downarrow\;\uparrow\rangle,\;|\uparrow\;\uparrow\rangle,\;|\downarrow\;\downarrow\rangle,\;|\uparrow\downarrow\,0\rangle,\;|0\,\uparrow\downarrow\rangle.
\end{align}
It is straightforward to diagonalize the Hamiltonian in this basis and the eigenstates are
\begin{align}\label{eigen_states}
|\psi_0\rangle&=\frac{4t}{a(c-U)}\Big(|\uparrow\;\downarrow\rangle-|\downarrow\;\uparrow\rangle\Big)+\frac{1}{a}\Big(|\uparrow\downarrow\,0\rangle+|0\,\uparrow\downarrow\rangle\Big),\notag\\
|\psi_1\rangle&=|\uparrow\;\uparrow\rangle,\notag\\
|\psi_2\rangle&=|\downarrow\;\downarrow\rangle,\notag\\
|\psi_3\rangle&=\frac{1}{\sqrt{2}}\Big(|\uparrow\;\downarrow\rangle+\;|\downarrow\;\uparrow\rangle\Big),\\
|\psi_4\rangle&=\frac{1}{\sqrt{2}}\Big(|\uparrow\downarrow\,0\rangle-|0\,\uparrow\downarrow\rangle\Big),\notag\\
|\psi_5\rangle&=\frac{4t}{b(c+U)}\Big(|\uparrow\;\downarrow\rangle-|\downarrow\;\uparrow\rangle\Big)-\frac{1}{b}\Big(|\uparrow\downarrow\,0\rangle+|0\,\uparrow\downarrow\rangle\Big),\notag
\end{align}
with
\begin{align}\label{abc}
a=\sqrt{\frac{32t^2}{(c-U)^2}+2},\qquad b=&\sqrt{\frac{32t^2}{(c+U)^2}+2},\notag\\
c=\sqrt{16t^2+U^2}&.
\end{align}
When $t\rightarrow0$, we obtain the Heitler-London solution where the second term of $|\psi_0\rangle$ vanishes and the two electrons are never localized at the same site. In the non-interacting limit where $U=0$, all the coefficients in $|\psi_0\rangle$ become equal and the state can easily be rewritten as a doubly occupied bonding $\sigma$-orbital, where the two electrons have a probability of $1/2$ for being localized on the same atom. In fact, any mean-field Hamiltonian without two-particle operators will yield such a ground state and it is a non-trivial task to produce the correct ground state energy in the dissociation limit using the non-interacting state as a reference for perturbation theory.

The associated eigenvalues are
\begin{align}\label{eigen_values}
E_0&=2\varepsilon_0+(U-c)/2,\\
E_1&=E_2=E_3=2\varepsilon_0,\\
E_4&=2\varepsilon_0+U,\\
E_5&=2\varepsilon_0+(U+c)/2.
\end{align}
In the present work we are interested in the ground state energy, which can be written as
\begin{align}
E^{Exact}&=2\varepsilon_0+\frac{U}{2}-\sqrt{4t^2+U^2/4}\label{E_exact}\\
&=2\varepsilon_0-2t+\frac{U}{2}-\frac{U^2}{16t}+\ldots,\qquad U\le4t,\label{E_U}\\
&=2\varepsilon_0-\frac{4t^2}{U}+\frac{t^4}{U^3}+\ldots,\qquad\qquad U\ge4t.\label{E_t}
\end{align}
From these expressions it is clear that it will be very challenging to derive an approximation for the ground state energy, which works well \textit{across coupling regimes}. Starting from the non-interacting system and calculating perturbative corrections in $U$ will systematically generate the perturbation series Eq. \eqref{E_U}, but it is hard to see how such an approach would reproduce the atomic limit Eq. \eqref{E_t}. In particular, higher order terms in the perturbation series would have to cancel the $U/2-2t$ terms, which are not present in Eq. \eqref{E_t}.

\subsection{The adiabatic connection}
In the present paper we calculate the ground state energy of the Hubbard dimer within various approximations. The calculations are based on the non-interacting reference state, but involve non-perturbative contributions to the energy. We define the correlation energy as the contribution to the energy beyond first order perturbation theory:
\begin{align}\label{E_c_def}
E_c\equiv E^{Exact}-2\varepsilon_0+2t-\frac{U}{2}.
\end{align}
This quantity will be calculated from various approximations to the density-density response function using the adiabatic connection and fluctuation-dissipation theorem.

To this end we introduce the $\lambda$-dependent Hamiltonian $H_\lambda$ by letting $U\rightarrow\lambda U$. The ground state of this Hamiltonian is written $|\psi_0\rangle_\lambda$ and becomes the true interacting ground state for $\lambda=1$ and the non-interacting ground state for $\lambda=0$. We then write the correlation energy as
\begin{align}
E_c=&\langle\psi_0^\lambda|H^\lambda|\psi_0^\lambda\rangle\Big|_{\lambda=1}-\langle\psi_0^\lambda|H^\lambda|\psi_0^\lambda\rangle\Big|_{\lambda=0}-\langle\psi_0^\lambda|V|\psi_0^\lambda\rangle\Big|_{\lambda=0}\notag\\
=&\int_0^1d\lambda\frac{d}{d\lambda}\langle\psi_0^\lambda|H^\lambda|\psi_0^\lambda\rangle-\langle\psi_0^{\lambda=0}|V|\psi_0^{\lambda=0}\rangle\\
=&\int_0^1d\lambda\langle\psi_0^\lambda|V|\psi_0^\lambda\rangle-\langle\psi_0^{\lambda=0}|V|\psi_0^{\lambda=0}\rangle,
\end{align}
where we applied the Hellman-Feynman theorem in the last line. We then use that
\begin{align}
\langle\psi_0^\lambda|V|\psi_0^\lambda\rangle=\frac{U}{2}\sum_{i\sigma\sigma'}\Big[\langle\psi_0^\lambda|n_{i\sigma}n_{i\sigma'}|\psi_0^\lambda\rangle-\langle\psi_0^\lambda|n_{i\sigma}|\psi_0^\lambda\rangle\delta_{\sigma\sigma'}\Big]\notag
\end{align}
and
\begin{align}\label{fluctuation}
\langle\psi_0^\lambda|n_{i\sigma}n_{i\sigma'}|\psi_0^\lambda\rangle&=n^\lambda_{i\sigma}n^\lambda_{i\sigma'}+\sum_{n\neq0}\langle\psi_0^\lambda|n_{i\sigma}|\psi_n^\lambda\rangle\langle\psi_n^\lambda|n_{i\sigma'}|\psi_0^\lambda\rangle\notag\\
&=n^\lambda_{i\sigma}n^\lambda_{i\sigma'}-\int_{-\infty}^\infty\frac{d\omega}{2\pi}\text{Im}\chi^\lambda_{ii,\sigma\sigma'}(\omega)
\end{align}
to get
\begin{align}
E_c&=-\frac{U}{2}\int_0^1d\lambda\int_{-\infty}^\infty\frac{d\omega}{2\pi}\sum_i\text{Im}\Big[\widetilde\chi_{ii}^\lambda(\omega)-\widetilde\chi_{ii}^0(\omega)\Big]\label{E_c}\\
&=-U\int_0^1d\lambda\int_{0}^\infty\frac{d\omega}{2\pi}\sum_i\Big[\widetilde\chi_{ii}^\lambda(i\omega)-\widetilde\chi_{ii}^0(i\omega)\Big]\label{E_c_im},
\end{align}
where we defined the spin-summed response function by:
\begin{align}
\widetilde\chi_{ij}^\lambda(\omega)=\sum_{\sigma\sigma'}\chi^\lambda_{ij,\sigma\sigma'}(\omega).
\end{align}
Here we used that the density $n_{i\sigma'}=\langle\psi_0^\lambda|n_{i\sigma}|\psi_0^\lambda\rangle=\frac{1}{2}$ is independent of $\lambda$. Note that the adiabatic connection in density functional theory involves an exchange-correlation potential \textit{defined} in such way that the density is independent of $\lambda$. In the present context, it is a convenient property of the model, which allow us to calculate total energies directly from the response function. In Eq. \eqref{E_c_im} we used that $\chi_{ii}(\omega)=\chi_{ii}(-\omega)$ and the fact that $\chi$ is analytic in the upper right quarter of the complex plane in order to change the integration to the positive imaginary axis. We note that the retarded and time-ordered response functions coincide on the positive imaginary axis and we will not distinguish between these as long as we consider imaginary frequencies. In the following it will often be convenient to write the response function in terms of imaginary frequencies, since then it is not necessary to keep track of the positive infinitesimals, which shift the poles away from the real axis. We can always restore the dependence on real frequencies by taking $\omega\rightarrow-i\omega$ and the hopping parameter $t\rightarrow t-i\eta$.

\section{Hedin's equations}\label{app_hedin}
To obtain an expression for $P$, we turn to Hedin's equations:\cite{hedin}
\begin{align}
G(12)&=G^0(12)+\int d(34)G^0(13)\Sigma(34)G(42)\label{G}\\
\Sigma(12)&=i\int d(34)G(13)\Gamma(324)W(41)\label{Sigma}\\
W(12)&=v(12)+\int d(34)v(13)P(34)W(42)\label{W}\\
P(12)&=-i\int d(34)G(13)G(41)\Gamma(342)\label{P}\\
\Gamma(123)&=\delta(12)\delta(13)\label{Gamma}\\
&\qquad+\int d(4567)\frac{\delta\Sigma(12)}{\delta G(45)}G(46)G(75)\Gamma(673)\notag,
\end{align}
where $G^0$ refers to the Green function associated with the Hartree Hamiltonian. Since the Hartree potential is just a constant in the model, we will not distinguish between Hartree and pure non-interacting Green functions in the following. All quantities here are time-ordered and the numbers denote combined space, time and spin indices such that $G^0(12)=G^0_{i_1i_2,\sigma_1\sigma_2}(t_1-t_2)$. In principle these equations should be iterated to self-consistency, however, for most practical applications some approximation is needed in order to proceed. For example, the GW approximation for the self-energy is obtained by neglecting the second term of Eq. \eqref{Gamma}, which leads to $\Sigma(12)=iG(12)W(21)$. This expression can now be used together with $G^0$ to obtain a one-shot expression for $G$ and $P$ ($G^0W^0$), or one can try to iterate the remaining equations to self-consistency (self-consistent $GW$). In the present work, we will impose a purely perturbative approach and calculate all quantities from $G^0$. We will therefore not need Eq. \eqref{G} and we replace $G$ in the remaining four equations by $G^0$, which is given by
\begin{align}
G^0(12)=-i\langle\psi^\lambda_0|Tc_{i_1\sigma_1}(t_1)c_{i_2\sigma_2}^\dag(t_2)|\psi^\lambda_0\rangle\Big|_{\lambda=0}.
\end{align}

\section{Hamiltonian formulation of the four-point Dyson equation}\label{app_hamilton}
In Section \ref{theory}, we evaluated the response function in various approximations by solving the Dyson equation directly. Due to the simplicity of the results it was straightforward to extract the poles, which correspond to excitation energies within a given approximation. An alternative approach is to express the response function in terms of eigenvalues and eigenstates of a particle-hole Hamiltonian. This is the approach used for the \textit{ab initio} calculations used in Section \ref{hydrogen} and has the advantage that it becomes much easier to expand a transition within a given approximation in terms of non-interacting transitions. With a static approximation for the four-point kernel \eqref{general_kernel} we can obtain the poles and transition matrix elements as eigenstates and eigenvalues of the Hamiltonian
\begin{align}\label{H_two}
H_{n_1n_2n_3n_4}=&(\varepsilon_{n_2}-\varepsilon_{n_1})\delta_{n_1n_3}\delta_{n_2n_4}\\
&+(f_{n_1}-f_{n_2})(U_{n_1n_2n_3n_4}-K_{n_1n_2n_3n_4}).\notag
\end{align}
Here $n_i$ labels single-particle spin orbitals with eigenvalue $\varepsilon_{n_1}$ and occupation $f_{n_i}$. In the case of  TDHF and BSE (but not in TDGW), the kernel $K$ is diagonal in spin in the sense that $K_{\sigma_1\sigma_2\sigma_3\sigma_4}=K_{\sigma_1\sigma_2}\delta_{\sigma_1\sigma_3}\delta_{\sigma_2\sigma_4}$ and the poles of the spin-summed response function can be obtained from the eigenvalues of the spin-summed Hamiltonian:
\begin{align}\label{H_two_summed}
H_{k_1k_2k_3k_4}&=(\varepsilon_{k_2}-\varepsilon_{k_1})\delta_{k_1k_3}\delta_{k_2k_4}\\
&+(f_{k_1}-f_{k_2})(U_{k_1k_2k_3k_4}-K_{k_1k_2k_3k_4}/2).\notag
\end{align}
Here $k_i$ labels single-particle orbitals with occupation factors $f_{k_i}$, which may be doubly occupied. In the present case of a Hubbard dimer $k_i$ may be either the bonding $\sigma$ orbital $(|1,0\rangle+|0,1\rangle)/\sqrt{2}$ or the anti-bonding $\sigma^*$ orbital $(|1,0\rangle-|0,1\rangle)/\sqrt{2}$. We then choose the ordered particle-hole basis $\{\sigma\otimes\sigma^*,\sigma^*\otimes\sigma,\sigma\otimes\sigma,\sigma^*\otimes\sigma^*\}$ and the site basis $\{11,12,21,22\}$. The $\sigma\otimes\sigma^*$ and $\sigma^*\otimes\sigma$ states can then be written as (1,-1,1,-1)/2 and (1,1,-1,-1)/2 respectively with respect to the site basis.

In general, the Hamiltonian will not be Hermitian, but can be shown to be \textit{pseudo-Hermitian}, which implies that the eigenvalues are either real or come in complex conjugated pairs. A pseudo-Hermitian matrix implies the existence of an inner product with respect to which the matrix is Hermitian and it can be shown that positive definiteness of this inner product implies a real spectrum of the matrix.\cite{ali1, ali2} In the present case, one can define an inner product as $\langle v|u\rangle_{\bar H}=\langle v|\bar H|u\rangle$, where $\bar H$ is given by\cite{gruning}
\begin{align}
\bar H=
\left[\begin{array}{cc}
I & O\\
O & -I
\end{array}\right]H.
\end{align}
It is straightforward to show that the Hamiltonians stated below are Hermitian with respect to the $\bar H$-inner product and the reality of the spectrum then depends on $\bar H$ being positive definite.

In the following we briefly state the two-particle Hamiltonians corresponding to spin-summed response function. It will be straightforward to verify that the Hamiltonians derived from RPA and TDHF always have a positive definite $\bar H$, whereas BSE and TDGW do not.

\subsection{RPA}
In this case the four-point kernel is not present and we just need the Coulomb interaction in an electron-hole basis. The four-point kernel is simply $U_{i_1i_2i_3i_4}=U\delta_{i_1i_2}\delta_{i_1i_3}\delta_{i_1i_4}$ and the Hamiltonian becomes
\begin{align}
H^{RPA}=
\left[\begin{array}{cc}
2t+U & U\\
-U & -2t-U
\end{array}\right].
\end{align}
Here we have neglected the $\{\sigma\otimes\sigma,\sigma^*\otimes\sigma^*\}$ sector, since all matrix elements involving these states vanish. The eigenvalues are $E^{RPA}_\pm=\pm\sqrt{4t^2+4tU}$ and in the atomic limit the eigenvectors become
\begin{align}
v_\pm\propto
\left(\begin{array}{c}
\pm2\sqrt{t/U}-1 \\
1
\end{array}\right).
\end{align}
Since the Hamiltonian is not Hermitian its eigenvectors are not orthogonal and in the atomic limit the two eigenvectors become parallel. The asymptotic state $(1,-1)$ can be written as $|12\rangle-|21\rangle$ in site basis and thus corresponds to a charge transfer excitation between atoms. The RPA thus correctly reproduces the vanishing matrix element of the density operator in the atomic limit.

\subsection{TDHF}
In the TDHF approximation the kernel is $K_{i_1i_2i_3i_4}=U\delta_{i_1i_2}\delta_{i_1i_3}\delta_{i_1i_4}$ and the Hamiltonian becomes
\begin{align}
H^{TDHF}=
\left[\begin{array}{cc}
2t+U/2 & U/2\\
-U/2 & -2t-U/2
\end{array}\right].
\end{align}
The results is thus vary similar to RPA, the only difference being that $U$ has been replaced by $U/2$ in the effective Hamiltonian. The eigenvalues are $E^{TDHF}_\pm=\pm\sqrt{4t^2+2tU}$ and as in the case of RPA, the eigenstates correspond to a charge transfer excitation in the atomic limit and therefore correctly reproduces the atomic limit.

\subsection{BSE}
In the BSE approximation the kernel is given by ($K_{i_1i_2i_3i_4}=W_{i_1i_2i_3i_4}$)
\begin{align}
\left[\begin{array}{cccc}
U-2U^2t/h^2 & 0 & 0 & 0\\
0 & 2U^2t/h^2 & 0 & 0\\
0 & 0 & 2U^2t/h^2 & 0\\
0 & 0 & 0 & U-2U^2t/h^2
\end{array}\right].
\end{align}
and the Hamiltonian becomes
\begin{align}
H^{BSE}=
\left[\begin{array}{cc}
2t+U/2 & U/2 + 2U^2t/h^2\\
-U/2 - 2U^2t/h^2 & -2t-U/2
\end{array}\right].
\end{align}
The eigenvalues are
\begin{align}
E^{BSE}_\pm=\pm\sqrt{4t^2+2tU-2U^3t/h^2-4U^4t^2/h^4},
\end{align}
which are equivalent to the expression \eqref{om_bse}. As in the case of RPA and TDHF the eigenstates are orthogonal in the non-interacting limit and becomes parallel and equal to $(1-,1)$ at the degeneracy point $t/U=(\sqrt{2}-1)/2$. When $t/U$ decreases beyond this point, the eigenvalues become imaginary and the eigenvectors are rotated with respect to each other in the complex plane. At the point where $t/U=0$ the eigenvectors become
\begin{align}
v_\pm=[e^{\pm\pi/6i}|\sigma\rangle\otimes|\sigma^*\rangle-e^{\mp\pi/6i}|\sigma^*\rangle\otimes|\sigma\rangle)]/\sqrt{2}.
\end{align}
Unlike RPA and TDHF, the eigenstates can not be written as a pure charge transfer excitation, when $t$ is decreased beyond the degeneracy point. However, expressing the eigenstates in terms of atomic orbitals, the part of the exciton wavefunction, which does not correspond to a charge transfer excitation becomes purely imaginary and does not contribute to the correlation energy.


\begin{thebibliography}{30}
\expandafter\ifx\csname natexlab\endcsname\relax\def\natexlab#1{#1}\fi
\expandafter\ifx\csname bibnamefont\endcsname\relax
  \def\bibnamefont#1{#1}\fi
\expandafter\ifx\csname bibfnamefont\endcsname\relax
  \def\bibfnamefont#1{#1}\fi
\expandafter\ifx\csname citenamefont\endcsname\relax
  \def\citenamefont#1{#1}\fi
\expandafter\ifx\csname url\endcsname\relax
  \def\url#1{\texttt{#1}}\fi
\expandafter\ifx\csname urlprefix\endcsname\relax\def\urlprefix{URL }\fi
\providecommand{\bibinfo}[2]{#2}
\providecommand{\eprint}[2][]{\url{#2}}

\bibitem[{\citenamefont{Helbig1 et~al.}(2009)\citenamefont{Helbig1, Tokatly,
  and Rubio}}]{helbig}
\bibinfo{author}{\bibfnamefont{N.}~\bibnamefont{Helbig}},
  \bibinfo{author}{\bibfnamefont{I.~V.} \bibnamefont{Tokatly}},
  \bibnamefont{and} \bibinfo{author}{\bibfnamefont{A.}~\bibnamefont{Rubio}},
  \bibinfo{journal}{J. Chem. Phys.} \textbf{\bibinfo{volume}{131}},
  \bibinfo{pages}{224105} (\bibinfo{year}{2009}).

\bibitem[{\citenamefont{Heitler and London}(1927)}]{heitler}
\bibinfo{author}{\bibfnamefont{W.}~\bibnamefont{Heitler}} \bibnamefont{and}
  \bibinfo{author}{\bibfnamefont{F.}~\bibnamefont{London}},
  \bibinfo{journal}{Zeitschrift für Physik} \textbf{\bibinfo{volume}{44}},
  \bibinfo{pages}{455} (\bibinfo{year}{1927}).

\bibitem[{\citenamefont{Olsen and Thygesen}(2013{\natexlab{a}})}]{olsen_rpa2}
\bibinfo{author}{\bibfnamefont{T.}~\bibnamefont{Olsen}} \bibnamefont{and}
  \bibinfo{author}{\bibfnamefont{K.~S.} \bibnamefont{Thygesen}},
  \bibinfo{journal}{Phys. Rev. B} \textbf{\bibinfo{volume}{87}},
  \bibinfo{pages}{075111} (\bibinfo{year}{2013}{\natexlab{a}}).

\bibitem[{\citenamefont{Furche}(2001)}]{furche}
\bibinfo{author}{\bibfnamefont{F.}~\bibnamefont{Furche}},
  \bibinfo{journal}{Phys. Rev. B} \textbf{\bibinfo{volume}{64}},
  \bibinfo{pages}{195120} (\bibinfo{year}{2001}).

\bibitem[{\citenamefont{Ren et~al.}(2012)\citenamefont{Ren, Rinke, Joas, and
  Scheffler}}]{ren_review}
\bibinfo{author}{\bibfnamefont{X.}~\bibnamefont{Ren}},
  \bibinfo{author}{\bibfnamefont{P.}~\bibnamefont{Rinke}},
  \bibinfo{author}{\bibfnamefont{C.}~\bibnamefont{Joas}}, \bibnamefont{and}
  \bibinfo{author}{\bibfnamefont{M.}~\bibnamefont{Scheffler}},
  \bibinfo{journal}{J. Mater. Sci} \textbf{\bibinfo{volume}{47}},
  \bibinfo{pages}{7447} (\bibinfo{year}{2012}).

\bibitem[{\citenamefont{He\ss{}elmann and G\"orling}(2010)}]{hesselmann1}
\bibinfo{author}{\bibfnamefont{A.}~\bibnamefont{He\ss{}elmann}}
  \bibnamefont{and}
  \bibinfo{author}{\bibfnamefont{A.}~\bibnamefont{G\"orling}},
  \bibinfo{journal}{Mol. Phys.} \textbf{\bibinfo{volume}{108}},
  \bibinfo{pages}{359} (\bibinfo{year}{2010}).

\bibitem[{\citenamefont{He\ss{}elmann and G\"orling}(2011)}]{hesselmann2}
\bibinfo{author}{\bibfnamefont{A.}~\bibnamefont{He\ss{}elmann}}
  \bibnamefont{and}
  \bibinfo{author}{\bibfnamefont{A.}~\bibnamefont{G\"orling}},
  \bibinfo{journal}{Phys. Rev. Lett.} \textbf{\bibinfo{volume}{106}},
  \bibinfo{pages}{093001} (\bibinfo{year}{2011}).

\bibitem[{\citenamefont{Olsen and Thygesen}(2012)}]{olsen_ralda1}
\bibinfo{author}{\bibfnamefont{T.}~\bibnamefont{Olsen}} \bibnamefont{and}
  \bibinfo{author}{\bibfnamefont{K.~S.} \bibnamefont{Thygesen}},
  \bibinfo{journal}{Phys. Rev. B} \textbf{\bibinfo{volume}{86}},
  \bibinfo{pages}{081103(R)} (\bibinfo{year}{2012}).

\bibitem[{\citenamefont{Olsen and Thygesen}(2013{\natexlab{b}})}]{olsen_ralda2}
\bibinfo{author}{\bibfnamefont{T.}~\bibnamefont{Olsen}} \bibnamefont{and}
  \bibinfo{author}{\bibfnamefont{K.~S.} \bibnamefont{Thygesen}},
  \bibinfo{journal}{Phys. Rev. B} \textbf{\bibinfo{volume}{88}},
  \bibinfo{pages}{115131} (\bibinfo{year}{2013}{\natexlab{b}}).

\bibitem[{\citenamefont{Gr{\"u}neis et~al.}(2009)\citenamefont{Gr{\"u}neis,
  Marsman, Harl, Schimka, and Kresse}}]{gruneis}
\bibinfo{author}{\bibfnamefont{A.}~\bibnamefont{Gr{\"u}neis}},
  \bibinfo{author}{\bibfnamefont{M.}~\bibnamefont{Marsman}},
  \bibinfo{author}{\bibfnamefont{J.}~\bibnamefont{Harl}},
  \bibinfo{author}{\bibfnamefont{L.}~\bibnamefont{Schimka}}, \bibnamefont{and}
  \bibinfo{author}{\bibfnamefont{G.}~\bibnamefont{Kresse}},
  \bibinfo{journal}{J. Chem. Phys.} \textbf{\bibinfo{volume}{131}},
  \bibinfo{pages}{154115} (\bibinfo{year}{2009}).

\bibitem[{\citenamefont{Henderson and Scuseria}(2010)}]{henderson}
\bibinfo{author}{\bibfnamefont{T.}~\bibnamefont{Henderson}} \bibnamefont{and}
  \bibinfo{author}{\bibfnamefont{G.}~\bibnamefont{Scuseria}},
  \bibinfo{journal}{Mol. Phys.} \textbf{\bibinfo{volume}{108}},
  \bibinfo{pages}{2511} (\bibinfo{year}{2010}).

\bibitem[{\citenamefont{Eshuis et~al.}(2012)\citenamefont{Eshuis, Bates, and
  Furche}}]{eshuis}
\bibinfo{author}{\bibfnamefont{H.}~\bibnamefont{Eshuis}},
  \bibinfo{author}{\bibfnamefont{J.~E.} \bibnamefont{Bates}}, \bibnamefont{and}
  \bibinfo{author}{\bibfnamefont{F.}~\bibnamefont{Furche}},
  \bibinfo{journal}{Theor. Chem. Acc.} \textbf{\bibinfo{volume}{131}},
  \bibinfo{pages}{1084} (\bibinfo{year}{2012}).

\bibitem[{\citenamefont{Caruso et~al.}(2013)\citenamefont{Caruso, Rohr,
  Hellgren, Ren, Rinke, Rubio, and Scheffler}}]{caruso}
\bibinfo{author}{\bibfnamefont{F.}~\bibnamefont{Caruso}},
  \bibinfo{author}{\bibfnamefont{D.~R.} \bibnamefont{Rohr}},
  \bibinfo{author}{\bibfnamefont{M.}~\bibnamefont{Hellgren}},
  \bibinfo{author}{\bibfnamefont{X.}~\bibnamefont{Ren}},
  \bibinfo{author}{\bibfnamefont{P.}~\bibnamefont{Rinke}},
  \bibinfo{author}{\bibfnamefont{A.}~\bibnamefont{Rubio}},
  \bibnamefont{and}
  \bibinfo{author}{\bibfnamefont{M.}~\bibnamefont{Scheffler}},
  \bibinfo{journal}{Phys. Rev. Lett.} \textbf{\bibinfo{volume}{110}},
  \bibinfo{pages}{146403} (\bibinfo{year}{2013}).

\bibitem[{\citenamefont{Aryasetiawan et~al.}(2002)\citenamefont{Aryasetiawan,
  Gunnarsson, and Rubio}}]{ferdi}
\bibinfo{author}{\bibfnamefont{F.}~\bibnamefont{Aryasetiawan}},
  \bibinfo{author}{\bibfnamefont{O.}~\bibnamefont{Gunnarsson}},
  \bibnamefont{and} \bibinfo{author}{\bibfnamefont{A.}~\bibnamefont{Rubio}},
  \bibinfo{journal}{Eurosphys. Lett.} \textbf{\bibinfo{volume}{57}},
  \bibinfo{pages}{683} (\bibinfo{year}{2002}).

\bibitem[{\citenamefont{Romaniello et~al.}(2009)\citenamefont{Romaniello,
  Guyot, and Reining}}]{romaniello1}
\bibinfo{author}{\bibfnamefont{P.}~\bibnamefont{Romaniello}},
  \bibinfo{author}{\bibfnamefont{G.}~\bibnamefont{Guyot}}, \bibnamefont{and}
  \bibinfo{author}{\bibfnamefont{L.}~\bibnamefont{Reining}},
  \bibinfo{journal}{J. Chem Phys.} \textbf{\bibinfo{volume}{131}},
  \bibinfo{pages}{154111} (\bibinfo{year}{2009}).

\bibitem[{\citenamefont{Romaniello et~al.}(2012)\citenamefont{Romaniello,
  Bechstedt, and Reining}}]{romaniello2}
\bibinfo{author}{\bibfnamefont{P.}~\bibnamefont{Romaniello}},
  \bibinfo{author}{\bibfnamefont{F.}~\bibnamefont{Bechstedt}},
  \bibnamefont{and} \bibinfo{author}{\bibfnamefont{L.}~\bibnamefont{Reining}},
  \bibinfo{journal}{Phys. Rev. B} \textbf{\bibinfo{volume}{85}},
  \bibinfo{pages}{155131} (\bibinfo{year}{2012}).

\bibitem[{\citenamefont{Kaasbjerg and Thygesen}(2010)}]{kaasbjerg}
\bibinfo{author}{\bibfnamefont{K.}~\bibnamefont{Kaasbjerg}} \bibnamefont{and}
  \bibinfo{author}{\bibfnamefont{K.~S.} \bibnamefont{Thygesen}},
  \bibinfo{journal}{Phys. Rev. B} \textbf{\bibinfo{volume}{81}},
  \bibinfo{pages}{085102} (\bibinfo{year}{2010}).

\bibitem[{\citenamefont{Rebolini et~al.}(2013)\citenamefont{Rebolini, Toulouse,
  and Savin}}]{rebolini}
\bibinfo{author}{\bibfnamefont{E.}~\bibnamefont{Rebolini}},
  \bibinfo{author}{\bibfnamefont{J.}~\bibnamefont{Toulouse}}, \bibnamefont{and}
  \bibinfo{author}{\bibfnamefont{A.}~\bibnamefont{Savin}},
  \bibinfo{journal}{arXiv:1304.1314}  (\bibinfo{year}{2013}).

\bibitem[{\citenamefont{Gritsenko et~al.}(2000)\citenamefont{Gritsenko, van
  Gisbergen, G{\"o}rling, and Baerends}}]{gritsenko}
\bibinfo{author}{\bibfnamefont{O.~V.} \bibnamefont{Gritsenko}},
  \bibinfo{author}{\bibfnamefont{A.~J.~A.} \bibnamefont{van Gisbergen}},
  \bibinfo{author}{\bibfnamefont{A.}~\bibnamefont{G{\"o}rling}},
  \bibnamefont{and} \bibinfo{author}{\bibfnamefont{E.~J.}
  \bibnamefont{Baerends}}, \bibinfo{journal}{J. Chem. Phys.}
  \textbf{\bibinfo{volume}{113}}, \bibinfo{pages}{8478} (\bibinfo{year}{2000}).

\bibitem[{\citenamefont{Enkovaara et~al.}(2010)}]{gpaw-paper}
\bibinfo{author}{\bibfnamefont{J.}~\bibnamefont{Enkovaara}}
  \bibnamefont{et~al.}, \bibinfo{journal}{J. Phys.: Condens. Matter}
  \textbf{\bibinfo{volume}{22}}, \bibinfo{pages}{253202}
  (\bibinfo{year}{2010}).

\bibitem[{\citenamefont{Yan et~al.}(2011)\citenamefont{Yan, Mortensen,
  Jacobsen, and Thygesen}}]{jun_response}
\bibinfo{author}{\bibfnamefont{J.}~\bibnamefont{Yan}},
  \bibinfo{author}{\bibfnamefont{J.~J.} \bibnamefont{Mortensen}},
  \bibinfo{author}{\bibfnamefont{K.~W.} \bibnamefont{Jacobsen}},
  \bibnamefont{and} \bibinfo{author}{\bibfnamefont{K.~S.}
  \bibnamefont{Thygesen}}, \bibinfo{journal}{Phys. Rev. B}
  \textbf{\bibinfo{volume}{83}}, \bibinfo{pages}{245122}
  (\bibinfo{year}{2011}).

\bibitem[{\citenamefont{Yan et~al.}(2012)\citenamefont{Yan, Jacobsen, and
  Thygesen}}]{jun_bse}
\bibinfo{author}{\bibfnamefont{J.}~\bibnamefont{Yan}},
  \bibinfo{author}{\bibfnamefont{K.~W.} \bibnamefont{Jacobsen}},
  \bibnamefont{and} \bibinfo{author}{\bibfnamefont{K.~S.}
  \bibnamefont{Thygesen}}, \bibinfo{journal}{Phys. Rev. B}
  \textbf{\bibinfo{volume}{86}}, \bibinfo{pages}{045208}
  (\bibinfo{year}{2012}).

\bibitem[{\citenamefont{Olsen et~al.}(2011)\citenamefont{Olsen, Yan, Mortensen,
  and Thygesen}}]{olsen_rpa1}
\bibinfo{author}{\bibfnamefont{T.}~\bibnamefont{Olsen}},
  \bibinfo{author}{\bibfnamefont{J.}~\bibnamefont{Yan}},
  \bibinfo{author}{\bibfnamefont{J.~J.} \bibnamefont{Mortensen}},
  \bibnamefont{and} \bibinfo{author}{\bibfnamefont{K.~S.}
  \bibnamefont{Thygesen}}, \bibinfo{journal}{Phys. Rev. Lett.}
  \textbf{\bibinfo{volume}{107}}, \bibinfo{pages}{156401}
  (\bibinfo{year}{2011}).

\bibitem[{\citenamefont{Yan et~al.}(2013)\citenamefont{Yan, Hummelsh\o{}j, and
  N\o{}rskov}}]{jun_rpa}
\bibinfo{author}{\bibfnamefont{J.}~\bibnamefont{Yan}},
  \bibinfo{author}{\bibfnamefont{J.~S.} \bibnamefont{Hummelsh\o{}j}},
  \bibnamefont{and} \bibinfo{author}{\bibfnamefont{J.~K.}
  \bibnamefont{N\o{}rskov}}, \bibinfo{journal}{Phys. Rev. B}
  \textbf{\bibinfo{volume}{87}}, \bibinfo{pages}{075207}
  (\bibinfo{year}{2013}).

\bibitem[{\citenamefont{Hedin}(1965)}]{hedin}
\bibinfo{author}{\bibfnamefont{L.}~\bibnamefont{Hedin}},
  \bibinfo{journal}{Phys. Rev.} \textbf{\bibinfo{volume}{139}},
  \bibinfo{pages}{A796} (\bibinfo{year}{1965}).

\bibitem[{\citenamefont{Wolniewicz}(1993)}]{wolnie}
\bibinfo{author}{\bibfnamefont{L.}~\bibnamefont{Wolniewicz}},
  \bibinfo{journal}{The Journal of Chemical Physics}
  \textbf{\bibinfo{volume}{99}}, \bibinfo{pages}{1851} (\bibinfo{year}{1993}).

\bibitem[{\citenamefont{Mott}(1949)}]{mott}
\bibinfo{author}{\bibfnamefont{N.~F.} \bibnamefont{Mott}},
  \bibinfo{journal}{Proc. Phys. Soc. A 62 416} \textbf{\bibinfo{volume}{62}},
  \bibinfo{pages}{416} (\bibinfo{year}{1949}).

\bibitem[{\citenamefont{Mostafazadeh}(2002{\natexlab{a}})}]{ali1}
\bibinfo{author}{\bibfnamefont{A.}~\bibnamefont{Mostafazadeh}},
  \bibinfo{journal}{Journal of Mathematical Physics}
  \textbf{\bibinfo{volume}{43}}, \bibinfo{pages}{205}
  (\bibinfo{year}{2002}{\natexlab{a}}).

\bibitem[{\citenamefont{Mostafazadeh}(2002{\natexlab{b}})}]{ali2}
\bibinfo{author}{\bibfnamefont{A.}~\bibnamefont{Mostafazadeh}},
  \bibinfo{journal}{Journal of Mathematical Physics}
  \textbf{\bibinfo{volume}{43}}, \bibinfo{pages}{2814}
  (\bibinfo{year}{2002}{\natexlab{b}}).

\bibitem[{\citenamefont{Gr{\"u}ning et~al.}(2009)\citenamefont{Gr{\"u}ning,
  Marini, and Gonze}}]{gruning}
\bibinfo{author}{\bibfnamefont{M.}~\bibnamefont{Gr{\"u}ning}},
  \bibinfo{author}{\bibfnamefont{A.}~\bibnamefont{Marini}}, \bibnamefont{and}
  \bibinfo{author}{\bibfnamefont{X.}~\bibnamefont{Gonze}},
  \bibinfo{journal}{Nano Lett.} \textbf{\bibinfo{volume}{9}},
  \bibinfo{pages}{2820} (\bibinfo{year}{2009}).

\end{thebibliography}

\end{document}